\documentclass[12pt]{article}
\usepackage{amssymb, amsmath, amsopn, amsthm}

\pdfoutput=1

\usepackage{bm,amsfonts,array,calc,rotating}
\usepackage{epsfig}
\usepackage{graphicx}
\usepackage{color}
\usepackage{slashed}
\usepackage[colorlinks=false]{hyperref}

\global\def\draftcontrol{0}

   \def\versionno{Higher Derivative Brane Couplings from String Amplitudes}

\catcode`\@=11

\expandafter\ifx\csname
draftcontrol\endcsname\relax\global\def\draftcontrol{0} \fi

{\count255=\time\divide\count255 by 60
\xdef\hourmin{\number\count255} \multiply\count255
by-60\advance\count255 by\time
\xdef\hourmin{\hourmin:\ifnum\count255<10 0\fi\the\count255}}
\def\draftdate{\number\month/\number\day/\number\year\ \ \ \hourmin }

\newcommand\makepapertitle{\par

  \begingroup
    \renewcommand\thefootnote{\@fnsymbol\c@footnote}%
    \def\@makefnmark{\rlap{\@textsuperscript{\normalfont\@thefnmark}}}%
    \long\def\@makefntext##1{\parindent 1em\noindent
            \hb@xt@1.8em{%
                \hss\@textsuperscript{\normalfont\@thefnmark}}##1}%
     \newpage
     \global\@topnum\z@   
     \@makepapertitle
     \thispagestyle{empty}\@thanks
  \endgroup
  \setcounter{footnote}{0}%
  \global\let\thanks\relax
  \global\let\makepapertitle\relax
  \global\let\@makepapertitle\relax
  \global\let\@thanks\@empty
  \global\let\@author\@empty
  \global\let\@date\@empty
  \global\let\@title\@empty
  \global\let\title\relax
  \global\let\author\relax
  \global\let\date\relax
  \global\let\and\relax
  \def\version{\let\version\@version\@gobble}
}
\def\@makepapertitle{%
  \newpage
   \ifnum\draftcontrol=1{}
   \version\versionno
   \vskip 5em%
   \else
   \hfill\hbox to 3cm {\parbox{4cm}{\@pubnum}\hss}%
   \vskip 5em%
   \fi
   \begin{center}%
   \let \footnote \thanks
      {\hskip -0\textwidth \hbox to 1\textwidth%
        {\centerline{\Large\bf{\noindent\@title}}}}%
     \vskip 2em%
     {\normalsize
       \lineskip .5em%
       \begin{tabular}[t]{c}%
         \@author
       \end{tabular}\par}%
     \vskip 1em%
     {\@bstract}%
     \end{center}%
     \vfill
     \@date%
     \vskip 1.5em%
     \noindent
     \rule{12em}{.02em}\par\noindent
     \@email%
   \par
}

\gdef\@pubnum{}
\def\pubnum#1{%
  \gdef\@pubnum{#1}}

\gdef\@bstract{}
\def\Abstract#1{%
  \gdef\@bstract{%
   \parbox{\textwidth-0pc}{%
   \centerline{\bf Abstract}\penalty1000
   \noindent
   \renewcommand\baselinestretch{1.0}
   {#1}}}
}

\gdef\@email{}
\def\email#1{%
   \gdef\@email{%
  {\small Email: {\tt #1}}}
}

\def\ps@paper{\let\@mkboth\@gobbletwo%
     \ifnum\draftcontrol=1
        \def\@oddfoot{\hbox to \textwidth{\tiny \versionno \hfil\tiny\draftdate}%
        \hskip -\textwidth \hbox to \textwidth{\hfil\rm\thepage\hfil}}%
     \else\def\@oddfoot{\hbox to \textwidth{\hfil\rm\thepage\hfil}}
     \fi
     \let\@evenfoot\@oddfoot
}

\def\body{\clearpage
          \pagestyle{paper}
        }
\newenvironment{acknowledgments}{%
\vskip 3.25ex
\noindent {\bf Acknowledgments}
}

\def\@version#1{\ifnum\draftcontrol=1
\typeout{}\typeout{#1}\typeout{}
\vskip3mm\centerline{\hbox{\fbox{\normalsize{\tt DRAFT -- #1 -- }
                   {\draftdate}}}}\vskip3mm
\fi}
\let\version\@version
\long\def\eqlabel#1{\ifnum\draftcontrol=1
                    \tag@false  
                    \tag*{(\theequation) \hbox to -0.2cm{\hspace{0cm}\small{#1}\hss}}
                    \refstepcounter{equation}
                    \edef\@currentlabel{\theequation}
                    \ltx@label{#1}          
                    \else
                    \label{#1}
                    \fi
                    }
\let\st@bibitem\@bibitem
\let\st@lbibitem\@lbibitem
\ifnum\draftcontrol=1
  \def\@bibitem#1{%
    \st@bibitem{#1}\a@@label{#1}\ignorespaces}
  \def\@lbibitem[#1]#2{%
    \st@lbibitem[#1]{#2}\a@@label{#2}\ignorespaces}
  \def\a@@label#1{%
    \gdef\a@lab{\smash{\normalfont\small#1}}
    \ifvmode
      \if@inlabel
        \global\setbox\@labels\hbox{%
          \llap{\a@lab\let\a@lab\relax
                \kern\@totalleftmargin\kern\marginparsep}%
          \box\@labels}%
      \fi
    \fi}
\fi

\renewcommand\baselinestretch{1.25}
\renewcommand\section{\@startsection {section}{1}{\z@}%
                                   {-3.5ex \@plus -1ex \@minus -.2ex}%
                                   {2.3ex \@plus.2ex}%
                                   {\normalfont\large\bfseries}}
\renewcommand\subsection{\@startsection{subsection}{2}{\z@}%
                                   {-3.25ex\@plus -1ex \@minus -.2ex}%
                                   {1.5ex \@plus .2ex}%
                                   {\normalfont\normalsize\bfseries}}
\renewcommand\subsubsection{\@startsection{subsubsection}{3}{\z@}%
                                   {-3.25ex\@plus -1ex \@minus -.2ex}%
                                   {1.5ex \@plus .2ex}%
                                   {\normalfont\normalsize\it}}
\renewcommand\paragraph{\@startsection{paragraph}{4}{\z@}%
                                   {-3.25ex\@plus -1ex \@minus -.2ex}%
                                   {1.5ex \@plus .2ex}%
                                   {\normalfont\normalsize\bf}}
\renewcommand\subparagraph{\@startsection{subparagraph}{5}{\z@}%
                                   {-1.25ex\@plus -1ex \@minus -.2ex}%
                                   {0ex \@plus .2ex}%
                                   {\normalfont\normalsize\it}}

\numberwithin{equation}{section}

\long\def\@makecaption#1#2{%
  \vskip\abovecaptionskip
  \sbox\@tempboxa{{\bf #1:} #2}%
  \ifdim \wd\@tempboxa >\hsize
    {\small\bf #1:} {\small #2}\par
  \else
    \global \@minipagefalse
    \hb@xt@\hsize{\hfil\box\@tempboxa\hfil}%
  \fi
  \vskip\belowcaptionskip}

\renewcommand*\l@section[2]{%
  \ifnum \c@tocdepth >\z@
    \addpenalty\@secpenalty
    \addvspace{.5em \@plus\p@}%
    \setlength\@tempdima{1.5em}%
    \begingroup
      \parindent \z@ \rightskip \@pnumwidth
      \parfillskip -\@pnumwidth
      \leavevmode \bfseries
      \advance\leftskip\@tempdima
      \hskip -\leftskip
      #1\nobreak\hfil \nobreak\hb@xt@\@pnumwidth{\hss #2}\par
    \endgroup
  \fi}
\renewcommand*\l@subsection{\addvspace{.0em \@plus\p@}\@dottedtocline{2}{1.5em}{2.3em}}
\renewcommand*\l@subsubsection{\addvspace{-.2em \@plus\p@}\@dottedtocline{3}{3.8em}{3.2em}}

\definecolor{refcol}{rgb}{0.2,0.2,0.8}
\definecolor{eqcol}{rgb}{.6,0,0}
\definecolor{purple}{cmyk}{0,1,0,0}

\gdef\@citecolor{refcol} \gdef\@linkcolor{eqcol}
\def\colorlinkspurple{\gdef\@urlcolor{purple}}
\def\colorlinksblue{\gdef\@urlcolor{blue}}
\def\colorlinksred{\gdef\@urlcolor{red}}

\def\revise#1       {\raisebox{-0em}{\rule{3pt}{1em}}%
                     \marginpar{\raisebox{.5em}{\vrule width3pt\
                     \vrule width0pt height 0pt depth0.5em
                     \hbox to 0cm{\hspace{0cm}{%
                     \parbox[t]{4em}{\raggedright\footnotesize{#1}}}\hss}}}}

\def\a{\alpha}

\def\d{\delta}

\def\k{\kappa}

\def\m{\mu}
\def\n{\nu}

\def\s{\sigma}

\def\G{\Gamma}

\def\P{\Pi}

\def\P{\Phi}

\def\p{\partial}

\def\p{\partial}

\def\inbar{\,\vrule height1.5ex width.4pt depth0pt}
\def\IC{\relax\hbox{$\inbar\kern-.3em{\rm C}$}}
\def\IN{\relax{\rm I\kern-.18em N}}
\def\IR{\relax{\rm I\kern-.18em R}}
\font\cmss=cmss10 \font\cmsss=cmss10 at 7pt
\def\IZ{\relax\ifmmode\mathchoice
{\hbox{\cmss Z\kern-.4em Z}}{\hbox{\cmss Z\kern-.4em Z}}
{\lower.9pt\hbox{\cmsss Z\kern-.4em Z}} {\lower1.2pt\hbox{\cmsss
Z\kern-.4em Z}}\else{\cmss Z\kern-.4em Z}\fi}

\def\half{\frac{1}{2}}
\def\a{\alpha}

\def\be{\begin{equation}}
\def\beq{\begin{eqnarray}}

\def\d{\delta}
\def\ee{\end{equation}}
\def\eeq{\end{eqnarray}}

\def\k{\kappa}

\def\m{\mu}
\def\n{\nu}

\def\om{\omega}

\def\s{\sigma}

\def\G{\Gamma}

\def\P{\Pi}

\def\P{\Phi}

\def\p{\partial}

\def\p{\partial}

\def\a{\alpha}
\def\G{\Gamma }

\def\e{\varepsilon}

\def\inbar{\,\vrule height1.5ex width.4pt depth0pt}

\def\half{\frac{1}{2}}

\def\half{\frac{1}{2}}

\def\zet          {{\mathbb Z}}

\def\ee           {{\it e}}

\def\tr           {{\rm tr}}
\def\Tr        {{\rm Tr}}

\def\sqr#1#2{{\vcenter{\vbox{\hrule height.#2pt
 \hbox{\vrule width.#2pt height#1pt \kern#1pt
 \vrule width.#2pt}\hrule height.#2pt}}}}

\newcommand{\C}[1]{$(\ref{#1})$}

\newcommand{\lp}{\left(}
\newcommand{\rp}{\right)}
\newcommand{\ls}{\left[}
\newcommand{\rs}{\right]}
\newcommand{\non}{\nonumber}
\newcommand{\hlf}{\frac{1}{2}}
\renewcommand{\be}{\begin{equation}}
\renewcommand{\ee}{\end{equation}}
\newcommand{\w}{\wedge}
\newcommand{\bea}{\begin{eqnarray}}
\newcommand{\eea}{\end{eqnarray}}

\newcommand{\wtb}{{\widetilde{b}}}
\newcommand{\wtc}{{\widetilde{c}}}
\newcommand{\wtpsi}{{\widetilde{\psi}}}
\newcommand{\wtS}{{\widetilde{S}}}
\newcommand{\wtphi}{{\widetilde{\phi}}}

\newcommand{\barz}{{\bar{z}}}

\newcommand{\barw}{{\bar{w}}}

\catcode`\@=12

\begin{document}

\pubnum{MIFPA-11-18}


\title{Disc amplitudes, picture changing and space-time actions}

\date{June 16, 2011}

\author{\\[.5cm]Katrin Becker, Guangyu Guo and Daniel Robbins\
\\[.2cm] \it Department of Physics, Texas A\&M University, \\ \it College Station, TX 77843, USA\\ [1.5cm]}

\Abstract{We study in detail the procedure for obtaining couplings of D-branes to closed string fields by evaluating string theory disc amplitudes.  We perform a careful construction of the relevant vertex operators and discuss the effects of inserting the boundary state which encodes the presence of the D-brane.  We confront the issue of non-decoupling of BRST-exact states and prove that the problem is evaded for the computations we need, thus demonstrating that our amplitudes are automatically gauge-invariant and independent of the distribution of picture charge.  Finally, we compute explicitly the two-point amplitudes of two NS-NS fields or one NS-NS and one R-R field on the disc, and we carefully compare all the lowest order terms with predictions from supergravity.}

 \email{kbecker, guangyu, robbins@physics.tamu.edu}

\makepapertitle

\body


\vskip 1em

\newpage

\section{Introduction}

As researchers have widened their exploration of the space of string theory vacua in recent years, more attention has been paid to compactifications which include D-branes and fluxes.  Much of the work being done focuses primarily on the effective theories which describe the low-energy dynamics of such theories, and so the effects of higher-derivative corrections to the string theory action are often ignored.  Caution is required however, as there are situations in which these contributions play a crucial role in establishing the consistency, or inconsistency, of solutions.

For example, consider the compactification of M-theory on a Calabi-Yau four-fold with flux.  The equation of motion for the three-form potential $C_3$ extended in the three space-time directions gives~\cite{Becker:1996gj} (assuming also supersymmetry of the solution)
\be
\label{eq:FourFoldLocalTadpole}
-d\ast_8df=\frac{1}{2}G_4\w\ast_8G_4+4\pi^2\sum_i\d^{(8)}(x-x_i)-4\pi^2X_8+\cdots,
\ee
where $f$ is a function related to the warp-factor, $x_i$ are the positions of space-filling M2-branes, and $X_8$ is a particular eight-derivative term built from four curvature tensors; other higher-derivative terms (eight-derivative and higher) are represented by $\cdots$.  Solving this equation locally for the warp-factor, the higher-derivative corrections are not relevant (assuming we are in the regime where the volume of the internal space is large in Planck units).  However, if we integrate (\ref{eq:FourFoldLocalTadpole}) over the internal space, then we find
\be
\label{eq:FourFoldGlobalTadpole}
0=\frac{1}{8\pi^2}\int\left|G_4\right|^2+N_{M2}-\frac{\chi}{24},
\ee
where $\chi$ is the Euler character of the four-fold, which comes from integrating $X_8$ over the internal space.  Thus, if we had naively ignored the higher-derivative correction, then we would have incorrectly concluded that a supersymmetric solution would require vanishing flux and no M2-branes.  Conversely, when we correctly include the higher-derivative corrections to the action, we conclude that fluxes (or branes) are a required ingredient on most Calabi-Yau four-folds ($\chi$ is typically positive on these spaces).  Thus, understanding of these higher-derivative terms was crucial for determining the correct consistency conditions on this class of solutions.

It is instructive to also consider another class of solutions that is related to the previous ones by duality, in which we compactify F-theory on the Calabi-Yau four-fold~\cite{Dasgupta:1999ss}.  In this case, the global consistency condition (\ref{eq:FourFoldGlobalTadpole}) corresponds, in IIB language, to a D3-brane tadpole and reads
\be
0=\int F_3\w H_3+N_{D3}-\frac{\chi}{24}.
\ee
From the point of view of IIB string theory, the contribution $\chi/24$ may at first appear mysterious, since, as in M-theory, the bulk action does not receive corrections until eight derivatives, but an eight derivative term would be too suppressed in the regime of a large, smooth compactification to give a topological contribution when integrated over the six-dimensional internal space of the IIB solution.  The resolution is that the IIB solution also necessarily includes D7-branes and O7-planes which wrap four-cycles of the internal space.  These brane actions can and do receive four-derivative corrections like~\cite{Green:1996dd,Cheung:1997az,Minasian:1997mm,Stefanski:1998yx,Craps:1998fn,Scrucca:1999uz}
\be
\label{eq:D7BraneTerms}
\d S_{D7}=-T_7\frac{\pi^2(\alpha')^2}{24}\int_{D7}C_4\w\lp\tr R_T\w R_T-\tr R_N\w R_N\rp,
\ee
and the integral of the $\tr R^2$ terms over the various seven brane world-volumes precisely reconstructs the contribution proportional to the Euler character of the four-fold.

The lesson from this example is that it is not only important to understand higher-derivative corrections in the bulk, but that higher-derivative corrections to D-brane actions can also play a pivotal role in determining the consistency of string compactifications; without taking these terms properly into account, we would reach mistaken conclusions about the space of valid constructions of string vacua.

However, terms like those in (\ref{eq:D7BraneTerms}) are not the full story; there are many other terms at the same order of derivatives which will appear in D-brane actions~\cite{Becker:2010ij}.  There are at least two routes by which we can learn about these additional terms; we can take the known terms such as those in (\ref{eq:D7BraneTerms}) and apply T-duality, or else we can try to compute the terms directly by evaluating scattering amplitudes.  
In the current work we will concentrate on the latter approach, and in fact we will largely be laying the ground-work for a more complete study by carefully examining many of the issues which arise when computing disc amplitudes and using them to reconstruct space-time actions.
In \cite{inprogress} we will use the tools presented in this paper to
present the gauge invariant completion of the four derivative corrections to the Wess-Zumino contribution to the D-brane action
found in ~\cite{Becker:2010ij}. These interactions have also
been considered by \cite{Garousi:2009dj,Garousi:2010ki,Garousi:2010bm,Garousi:2011ut,Garousi:2011fc}.

To compute the terms of interest, we must evaluate scattering amplitudes in which various closed string fields interact with a D-brane\footnote{It is also interesting to compute the scattering of closed string fields from orientifold planes, by computing string amplitudes with a crosscap instead of a boundary.  It is well known how to accommodate that situation in the boundary state formalism.  We don't work out the details in the present paper, but expect that most of our techniques and results carry over to that case easily.}.  We will restrict ourselves to tree-level computations, so the relevant amplitudes are given by insertions of multiple closed string vertex operators on a world-sheet with the topology of a disc.  We will study this problem using the boundary state formalism~\cite{Callan:1986bc,Kostelecky:1987px}.  In this formalism we work with the usual vertex operators and BRST cohomology that we would use on the sphere~\cite{Friedan:1985ge}, but to account for the effect of a world-sheet boundary, we insert a boundary state $|B\rangle$ which encodes the boundary conditions of fields in the presence of the D-brane.  We also need to include a propagator which pushes this induced boundary out to the first closed string insertion point, and a ghost factor $b_0+\widetilde{b}_0$.

Though the boundary state itself is annihilated by the total (left- plus right-moving) BRST charge, the extra ghost insertion is not invariant, and this fact leads to many subtle issues which do not occur for sphere amplitudes.  For example, it is not necessarily true that BRST-exact operators decouple from disc amplitudes, and this potentially leads to disturbing consequences.  Gauge transformations of the space-time fields are represented by shifting the corresponding vertex operators by BRST-exact pieces, so if these do not decouple it would mean that the scattering amplitude was not gauge invariant, which should not happen for physical quantities.  A related issue is that amplitudes in the NSR formalism are not supposed to depend upon how the total picture charge is distributed among the various operators, but verifying this property typically relies on the decoupling of certain BRST-exact states.  So, in order to do a careful analysis of disc amplitudes with closed string insertions, it is important to really understand these issues and whether they affect the integrity of our answers.

The outline of this paper is as follows.  In section \ref{sec:VertexOperators} we start by constructing the physical state vertex operators by computing the relevant BRST-cohomologies to describe the massless fields of the superstring.  This section also serves as a summary of many of our conventions for OPEs that we will need when we proceed to compute amplitudes.  In section \ref{sec:BoundaryState}, we discuss the boundary state $|B\rangle$ and its effects on the computation, and in particular we demonstrate how we can use $|B\rangle$ to convert all right-moving fields in the computation into left-movers to facilitate the evaluation of the amplitude.  Section \ref{sec:BRSTExact} deals with BRST-exact states in the amplitude and shows that they can give rise to boundary terms that need not vanish.  However, by appealing to the analyticity of the amplitude as a function of the external momenta, we demonstrate that in a broad variety of circumstances (we also discuss the situations where this argument fails) the boundary terms do vanish identically.  We show that this proves that the amplitudes are indeed gauge invariant and that the result is independent of how we distribute the total picture charge.  Most of this discussion focusses on the two-point functions for simplicity.  Next, in section \ref{sec:TwoPointAmplitudes} we explicitly compute various two-point functions on the disc, and in section \ref{sec:ComparisonToSUGRA} we compare the leading terms in the momentum expansion with predictions from supergravity and show exact agreement.  

\section{Vertex operators}
\label{sec:VertexOperators}

\subsection{Notation and conventions}

We start with the usual matter and ghosts on the world-sheet with OPEs on the complex plane\footnote{The OPE for $\psi^\m$ differs from~\cite{Becker:2010ij} by a sign; we have changed conventions to match most of the literature on boundary states.}
\bea
\label{eq:PlaneCorrelators}
X^\m(z,\bar{z})X^\n(w,\bar{w}) &\sim& -\eta^{\m\n}\ln\left|z-w\right|^2,\non\\
\psi^\m(z)\psi^\n(w) &\sim& \frac{\eta^{\m\n}}{z-w},\non\\
b(z)c(w)\sim c(z)b(w) &\sim& \frac{1}{z-w},\\
\phi(z)\phi(w) &\sim& -\ln\lp z-w\rp,\non\\
\eta(z)\xi(w)\sim\xi(z)\eta(w) &\sim& \frac{1}{z-w},\non
\eea
and similarly for the anti-holomorphic fields.

The (holomorphic) ghost charge $q_g$ and picture charge $q_P$ of an operator $\mathcal{O}$ are given by\footnote{These definitions are not universally agreed upon.  Our choice of picture charge is chosen so that it commutes with BRST charge, $[Q_P,Q_{BRST}]=0$.  Our ghost charge satisfies $[Q_g,Q_{BRST}]=Q_{BRST}$, i.e. the BRST current has ghost charge one, but this does not determine it uniquely; we could add any multiple of $J_P$ to $J_g$.  The precise form here is fixed by also requiring that the picture changing operators have ghost number zero, so that they relate states or operators with the same ghost number.}
\bea
\ls Q_g,\mathcal{O}\rs(0)=\oint\frac{dz}{2\pi i}J_g(z)\mathcal{O}(0) &=& q_g\mathcal{O}(0),\\
\ls Q_P,\mathcal{O}\rs(0)=\oint\frac{dz}{2\pi i}J_P(z)\mathcal{O}(0) &=& q_P\mathcal{O}(0),
\eea
where
\be
J_g=:cb:+:\eta\xi:,\qquad J_P=-\p\phi+:\xi\eta:,
\ee
and $Q_g$ and $Q_P$ are the corresponding charge operators.  The charges and conformal weights of several of these fields are listed in Table \ref{table:nonlin}.

\begin{table}[htp!]
\centering
\begin{tabular}{c c c c} 
\hline 
& $q_g$ & $q_P$ & $h$ \\ [0.5ex] 
\hline 
$c$ & 1 & 0 & -1 \\ 
$b$ & -1 & 0 & 2 \\
$\eta$ & 1 & -1 & 1 \\
$\xi$ & -1 & 1 & 0 \\
$e^{n \phi}$  & 0 & $n$ & $-n \left(n +2 \right)/2 $ \\ [1ex] 
$\partial X^\m$ & 0 & 0 & 1 \\
$e^{i k\cdot X}$ & 0 & 0 & $k^2/2$ \\
$\psi^\mu $ & 0 & 0 & $1/2 $\\
\hline 
\end{tabular}
\label{table:nonlin} 
\caption{Left-moving fields with their ghost charge, picture charge, and conformal weight}
\end{table}

Finally, in our conventions the left-moving BRST charge is given by
\be
\ls Q_{BRST},\mathcal{O}\rs(0)=\oint\frac{dz}{2\pi i}J_{BRST}(z)\mathcal{O}(0),
\ee
where
\be
J_{BRST}=J_0+J_1+J_2,
\ee
\bea
J_0 &=& c\lp -\hlf\p X^\m\p X_\m-\hlf\psi^\m\p\psi_\m-\hlf\p\phi\p\phi-\p^2\phi-\eta\p\xi+\p cb\rp,\\
J_1 &=& -\hlf e^\phi\eta\psi^\m\p X_\m,\\
J_2 &=& \frac{1}{4}e^{2\phi}b\eta\p\eta.
\eea

\subsection{Physical States}
\label{subsec:PhysicalStates}

In the closed string we have both holomorphic (left-moving) fields and their anti-holomorphic (right-moving) counterparts, which we denote throughout the paper with tildes.

Physical states in the closed string correspond~\cite{Vafa:1987es,AlvarezGaume:1988bg,Zwiebach:1988qp,Nelson:1988ic,La:1989xk,Distler:1990ea} to classes in the {\it{semirelative BRST-cohomology}}, i.e. states which are annihilated by the operator $b_0^-=b_0-\wtb_0$ and by the total BRST charge $Q=Q_{BRST}+\widetilde{Q}_{BRST}$, modulo states which can be written as $Q$ acting on something that is annihilated by $b_0^-$.  In terms of operators, we need
\be
\ls b_0^-,V\rs=\ls Q,V\rs=0,\qquad\d V=\ls Q,U\rs,\quad\ls b_0^-,U\rs=0.
\ee

In practice, obtaining a complete characterization of this cohomology is difficult, especially since the $b_0^-$ condition does not factorize between the left- and right-movers.  In this section we will collect various results in the literature to explain how we have a bit more flexibility when states have well-defined non-vanishing momentum, $p^\m\ne 0$, and when we only intend to insert our vertex operators into disc amplitudes, and are not worried about arbitrary higher-genus Riemann surfaces.

\subsubsection{Chiral states}

Let us start with states that are purely left-moving.  In this case there are two cohomologies we can define, either that given by BRST-closed states modulo BRST-exact states, without further restrictions, or the relative cohomology of states which are annihilated by $b_0$.  Note that the BRST charge has picture charge zero, ghost charge one, conformal weight zero, and zero momentum,
\be
\label{eq:BRSTQuantumNumbers}
\ls Q_P,Q_{BRST}\rs=\ls L_0,Q_{BRST}\rs=\ls\widehat{p}^\m, Q_{BRST}\rs=0,\qquad\ls Q_g,Q_{BRST}\rs=Q_{BRST},
\ee
where
\be
\ls\widehat{p}^\m,\mathcal{O}\rs(0)=\oint\frac{dz}{2\pi i}\p X^\m(z)\mathcal{O}(0),
\ee
is the momentum operator.  Note that we also have
\be
\label{eq:b0QuantumNumbers}
\ls Q_P,b_0\rs=\ls L_0,b_0\rs=\ls\widehat{p}^\m,b_0\rs=0,\qquad\ls Q_g,b_0\rs=-b_0,
\ee
as well as
\be
\label{eq:MutuallyCommutingOperators}
\ls Q_P,Q_g\rs=\ls Q_P,L_0\rs=\ls Q_P,\widehat{p}^\m\rs=\ls Q_g,L_0\rs=\ls Q_g,\widehat{p}^\m\rs=\ls L_0,\widehat{p}^\m\rs=0.
\ee

Now from (\ref{eq:BRSTQuantumNumbers}), (\ref{eq:b0QuantumNumbers}), and (\ref{eq:MutuallyCommutingOperators}), we see that without loss of generality we can restrict attention to particular eigenspaces of $Q_P$, $L_0$, and $\widehat{p}^\m$, and we can grade the cohomology by ghost number.  In other words, we can define spaces
\be
C^n_{P,\lambda,p^\m}=\left\{V|\ls Q_P,V\rs=PV,\ \ls L_0,V\rs=\lambda V,\ \ls\widehat{p}^\m,V\rs= p^\m V,\ \ls Q_g,V\rs=nV\right\},
\ee
and then define the absolute and relative chiral cohomologies,
\be
H^n_{P,\lambda,p^\m}=\frac{\left\{V\in C^n_{P,\lambda,p^\m}|\ls Q_{BRST},V\rs=0\right\}}{\left\{\ls Q_{BRST},U\rs|U\in C^{n-1}_{P,\lambda,p^\m}\right\}},
\ee
and
\be
H^n_{R;P,p^\m}=\frac{\left\{V\in C^n_{P,0,p^\m}|\ls b_0,V\rs=\ls Q_{BRST},V\rs=0\right\}}{\left\{\ls Q_{BRST},U\rs|U\in C^{n-1}_{P,0,p^\m},\ \ls b_0,U\rs=0\right\}}.
\ee


Note that in the relative chiral cohomology we can drop the $L_0$ eigenvalue because $\ls b_0,V\rs=\ls Q_{BRST},V\rs=0$ implies that $\ls L_0,V\rs=0$.  In the case of the absolute chiral cohomology, for $\lambda\ne 0$ we have for BRST-closed operators $V$,
\be
V=\ls Q_{BRST},\lp\lambda^{-1}\ls b_0,V\rs\rp\rs,
\ee
implying that $H^n_{P,\lambda,p^\m}=0$ for $\lambda\ne 0$.  So there too we will assume that $\lambda=0$.  We will assume, in both cohomologies, that we have fixed $P$ and $p^\m$ to some agreed upon values, and we will not bother to include them as subscripts.  Thus we will talk about the {\it{absolute chiral cohomology}} $H^n$ and the {\it{relative chiral cohomology}} $H_R^n$.

These two cohomologies fit together into a long exact sequence (see for example~\cite{Witten:1992yj}),
\be
\label{eq:ChiralLES}
\cdots\longrightarrow H_R^n\stackrel{i}{\longrightarrow}H^n\stackrel{b_0}{\longrightarrow}H_R^{n-1}\stackrel{\left\{Q,c_0\right\}}{\longrightarrow}H_R^{n+1}\stackrel{i}{\longrightarrow}H^{n+1}\longrightarrow\cdots,
\ee
where the map $i$ is simply inclusion, and the other maps indicate taking the commutator with $b_0$ or with $\left\{Q_{BRST},c_0\right\}$ respectively.  It is easy to check that the kernel of each map is the image of the previous map.

Now we mention some results from the literature.  It is possible to construct picture changing operators
\be
X(z)=\left\{Q_{BRST},2\xi(z)\right\},\qquad\mathrm{and}\qquad Y(z)=-2:c\p\xi e^{-2\phi}(z):,
\ee
whose zero mode pieces $X_0$ and $Y_0$ commute with $Q_{BRST}$, $Q_g$, $L_0$, and $\widehat{p}^\m$, and which carry picture charge $+1$ and $-1$ respectively, and which further satisfy
\be
X_0Y_0=Y_0X_0=1+\left\{Q_{BRST},\cdots\right\}.
\ee
Thus, these operators can be used to construct an isomorphism between the absolute cohomology with picture $P$ and the one with picture $P+k$ for any $k\in\zet$.

Unfortunately, $Y_0$ does not commute with $b_0$, so these operators cannot be used to construct an isomorphism of relative cohomologies.  However, for $p^\m\ne 0$, it was shown by Berkovits and Zwiebach~\cite{Berkovits:1997mc} that one can construct an alternative operator 
\be
Y'(z)=-2\ell_\m :e^{-\phi}\psi^\m(z):,\qquad\mathrm{where}\qquad\ell^\m p_\m=1,
\ee
whose zero mode piece $Y_0'$ does commute with $b_0$ as well as $Q_{BRST}$, $Q_g$, $L_0$, and $\widehat{p}^\m$, and which, when restricted to the $p^\m\ne 0$ eigenspace satisfies
\be
X_0Y_0'=Y_0'X_0=1+\left\{Q_{BRST},\cdots\right\},
\ee
where $\cdots$ is annihilated by $b_0$.  These then establish isomorphisms between $H_R^n$ at picture $P$ and $H_R^n$ at picture $P+k$, $k\in\zet$.

Next, we have a result due to Lian and Zuckerman~\cite{Lian:1989cy}, generalizing~\cite{Frenkel:1986dg} for the bosonic string, which shows that, again for $p^\m\ne 0$, $H_R^n=0$ unless $n=1$.  In fact, looking carefully at their results, they prove this only for pictures $-1$, $-1/2$, and $-3/2$ (they start with a vacuum in one of these pictures and consider only states built by acting on the vacuum with a finite number of $\beta$ and $\gamma$ superghost oscillators), but combining this result with the isomorphisms constructed by~\cite{Berkovits:1997mc}, we have that $H_R^n=0$ for $n\ne 1$ in any picture and any non-zero $p^\m$.

If we plug the Lian-Zuckerman vanishing result into the sequence (\ref{eq:ChiralLES}), we learn that for $p^\m\ne 0$, the maps
\be
H_R^1\stackrel{i}{\longrightarrow}H^1,\qquad H^2\stackrel{b_0}{\longrightarrow}H_R^1,
\ee
are isomorphisms, and that $H^n=0$ if $n\ne 1,2$.  The first isomorphism in particular implies that for any BRST-closed operator $V$ of ghost charge one, there exist operators $W$ and $U$ satisfying
\be
\ls b_0,W\rs=\ls b_0,U\rs=0,\qquad\ls Q_{BRST},W\rs=0,
\ee
such that
\be
V=W+\ls Q_{BRST},U\rs.
\ee

\subsubsection{Closed string states}

In the closed string there are three different cohomologies that arise, the {\it{absolute cohomology}} $\mathcal{H}^n$, the {\it{semirelative cohomology}} $\mathcal{H}_S^n$, and the {\it{relative cohomology}}, $\mathcal{H}_R^n$, defined by
\bea
\mathcal{H}^n &=& \frac{\left\{V|\ls Q,V\rs=0,\ \ls Q_G,V\rs=nV\right\}}{\left\{\ls Q,U\rs|\ls Q_G,U\rs=(n-1)U\right\}},\\
\mathcal{H}_S^n &=& \frac{\left\{V|\ls Q,V\rs=\ls b_0^-,V\rs=0,\ \ls Q_G,V\rs=nV\right\}}{\left\{\ls Q,U\rs|\ls b_0^-,U\rs=0,\ \ls Q_G,U\rs=(n-1)U\right\}},\\
\mathcal{H}_R^n &=& \frac{\left\{V|\ls Q,V\rs=\ls b_0,V\rs=\ls \wtb_0,V\rs=0,\ \ls Q_G,V\rs=nV\right\}}{\left\{\ls Q,U\rs|\ls b_0,U\rs=\ls\wtb_0,U\rs=0,\ \ls Q_G,U\rs=(n-1)U\right\}},
\eea
where we recall that $Q=Q_{BRST}+\widetilde{Q}_{BRST}$, $b_0^-=b_0-\wtb_0$, and we define the total ghost charge, $Q_G=Q_g+\widetilde{Q}_g$.  We assume that we are working at fixed left and right pictures $P$ and $\widetilde{P}$ and fixed momentum $p^\m$, and vanishing eigenvalues for $L_0$ and $\widetilde{L}_0$.

With standard techniques it is easy to express $\mathcal{H}^n$ and $\mathcal{H}_R^n$ in terms of chiral cohomologies by K\"unneth formulae,
\be
\label{eq:Kunneth}
\mathcal{H}^n=\sum_{k+\ell=n}H^k\otimes\widetilde{H}^\ell,\qquad\mathcal{H}_R^n=\sum_{k+\ell=n}H_R^k\otimes\widetilde{H}_R^\ell.
\ee
Using the vanishing theorems of the previous subsection, this implies that for $p^\m\ne 0$, we have
\be
\mathcal{H}^n=0,\quad\mathrm{for}\ n\ne 2,3,4,\qquad\mathrm{and}\qquad\mathcal{H}_R^n=0,\quad\mathrm{for}\ n\ne 2.
\ee

Because the semirelative condition does not factorize between left and right, we can't use such a simple decomposition for the semirelative complex, which are the states of legitimate physical interest.  However, there are long exact sequences~\cite{Witten:1992yj} anologous to the chiral (\ref{eq:ChiralLES}),
\be
\cdots\longrightarrow\mathcal{H}_S^n\stackrel{i}{\longrightarrow}\mathcal{H}^n\stackrel{b_0^-}{\longrightarrow}\mathcal{H}_S^{n-1}\stackrel{\{Q,c_0^-\}}{\longrightarrow}\mathcal{H}_S^{n+1}\stackrel{i}{\longrightarrow}\mathcal{H}^{n+1}\longrightarrow\cdots,
\ee
and
\be
\cdots\longrightarrow\mathcal{H}_R^n\stackrel{i}{\longrightarrow}\mathcal{H}_S^n\stackrel{b_0^+}{\longrightarrow}\mathcal{H}_R^{n-1}\stackrel{\{Q,c_0^+\}}{\longrightarrow}\mathcal{H}_R^{n+1}\stackrel{i}{\longrightarrow}\mathcal{H}_S^{n+1}\longrightarrow\cdots.
\ee
Plugging in our vanishing theorems we learn that, for $p^\m\ne 0$, there are isomorphisms
\be
\label{eq:ClosedStringIsomorphisms}
\mathcal{H}^4\cong\mathcal{H}_S^3\cong\mathcal{H}_R^2\cong\mathcal{H}_S^2\cong\mathcal{H}^2,
\ee
and the only other nonvanishing cohmology group is $\mathcal{H}^3$ which can be obtained from (\ref{eq:Kunneth}) or from the short exact sequence
\be
0\longrightarrow\mathcal{H}_S^3\stackrel{i}{\longrightarrow}\mathcal{H}^3\stackrel{b_0^-}{\longrightarrow}\mathcal{H}_S^2\longrightarrow 0.
\ee

To summarize, this shows that if our purpose is simply to find the spectrum of physical closed string states at non-zero momentum we have many choices, since there are many cohomology groups which are isomorphic to the desired\footnote{For unintegrated vertex operators, as we are discussing here, we are almost always interested in total ghost number two.} $\mathcal{H}_S^2$.  

However, when we wish to insert our vertex operators into amplitudes then we need to be more careful.  The semirelative condition is imposed in order to make correlation functions well-defined on arbitrary higher genus Riemann surfaces~\cite{Vafa:1987es,AlvarezGaume:1988bg,Nelson:1988ic,La:1989xk,Distler:1990ea}.  In particular, if $U$ corresponds to a state that is not annihilated by $b_0^-$, then the BRST exact insertion $[Q,U]$ is not guaranteed to decouple from amplitudes.  Thus if $V$ represents a class in $\mathcal{H}^2$ with $p^\m\ne 0$, then the last isomorphism in (\ref{eq:ClosedStringIsomorphisms}) guarantees that we can write
\be
V=W+\ls Q,U\rs,\qquad\ls b_0^-,W\rs=0,
\ee
but the second term may not decouple and so they may give different results inside correlation functions.  However, we will argue in section \ref{sec:BRSTExact} that for correlation functions on the disc at generic nonzero momenta, that $[Q,U]$ will in fact decouple, even if $U$ is not annihilated by $b_0^-$.

Our purpose in clarifying all these issues is that below we will find R-R operators in $\mathcal{H}^2$ which are not in $\mathcal{H}_S^2$ but which enjoy certain desirable properties in the context of disc amplitudes with one R-R field and several NS-NS fields.  Using $\mathcal{H}_S^2$, one finds that in the $(-1/2,-1/2)$ picture one can write R-R vertex operators which are manifestly gauge invariant (they depend on the R-R field strength $F$ rather than the potential $C$), but we must put at least one of the NS-NS fields in an asymmetric picture and we lose the manifest exchange symmetry between the NS-NS fields.  Alternatively in the $(-3/2,-1/2)$ picture we can maintain the exchange symmetry, but we lose the gauge invariance and in fact must work in a somewhat awkward gauge.  However, if we are willing to relax the semirelative condition and work with R-R operators in $\mathcal{H}^2$, then we can find something which combines both of the desirable properties.

\subsection{Vertex operators}


\subsubsection{Open string}

Whether we wish to compute the absolute chiral cohomology at ghost number one, $H^1$, or the relative chiral cohomology $H_R^1$, we start the same way.  At a given mass level, say $p^2=0$, and a given choice of picture, we can classify all possible open-string vertex operators that have conformal weight zero and ghost number one and are constructed out of the basic fields (these fields and their charges and weights were summarized in Table \ref{table:nonlin}).  Suppose we want operators of picture $P$.  If we start with a contribution $e^{(n+P)\phi}$, then to get the correct picture charge we must also include either $n$ $\eta$'s, if $n>0$, or $|n|$ $\xi$'s if $n\le 0$.  Then to get the correct ghost number we need either $n-1$ $b$'s for $n>0$ or $1-n$ $c$'s for $n\le 0$.  Calculating the conformal weight in either case, we find
\bea
\ls b\p b\cdots\p^{n-2}b\eta\p\eta\cdots\p^{n-1}\eta e^{(n+P)\phi}\rs=\frac{n^2-2Pn-\lp P^2+2P+2\rp}{2}, &\quad& n>0,\non\\
\ls c\p c\cdots \p^{|n|}c\p\xi\p^2\xi\cdots\p^{|n|}\xi e^{(n+P)\phi}\rs=\frac{n^2-2\lp P+1\rp n-\lp P^2+2P+2\rp}{2}, &\quad& n\le 0.\non
\eea
We can then add more pieces which don't change the picture charge or ghost number, including contributions from the matter sector, additional $\eta$-$\xi$ or $b$-$c$ pairs, or derivatives acting on any of these fields.  All these contributions, however, will only increase the conformal weight, so we need the basic contribution above to have weight less than or equal to zero.  For instance, if we want operators of picture $P=-1$, we have either $(n^2+2n-1)/2$ for $n>0$, or $(n^2-1)/2$ for $n\le 0$.  The only viable solutions are then $n=0$ or $n=-1$, and the corresponding possible operators are
\be
V_{-1}=\ls\a_\m ce^{-\phi}\psi^\m+\beta c\p c\p\xi e^{-2\phi}\rs e^{ipX}.
\ee
Imposing the condition that this is BRST closed, we find that $\beta=0$ and $p^\m\a_\m=0$.  Note that if we imposed $[b_0,V]=0$ first, we would set $\beta=0$ before considering BRST closure, but the end result is the same.

In general, under gauge transformations vertex operators change by BRST exact operators
\begin{equation}
\delta V_P = \left[ Q_{BRST},U_P \right],
\end{equation}
where $U_P$ has the same momentum and picture charge as $V_P$, and vanishing conformal weight and ghost charge.

Similar considerations to those above then allow us to classify all possible gauge transformations.  For picture $-1$, for example, we have only
\be
U_{-1}=i\lambda c\p\xi e^{-2\phi}e^{ipX},
\ee
generating the gauge transformations
$
\d\a_\m=\hlf\lambda p_\m.
$

Similar calculations give, for picture $0$,
\be
V_0=\a_\m\ls c\lp\p X^\m-ip_\n\psi^\n\psi^\m\rp-\hlf e^\phi\eta\psi^\m\rs e^{ipX},
\ee
subject to $p^2=0$ and $p^\m\a_\m=0$.
In this case
\begin{equation}
U_0=-i\lambda e^{ipX}
\end{equation}
corresponding to the gauge transformations $\d\a_\m=\lambda p_\m$.

In both these pictures, the absolute and relative cohomologies are identical.

For picture $-\frac{3}{2}$, in the absolute cohomology $H^1$ we have
\be
V_{-\frac{3}{2}}=\ls \a_Ac\p c\p\xi e^{-\frac{5}{2}\phi}+i\beta_Ace^{-\frac{3}{2}\phi}\rs S^Ae^{ipX},
\ee
subject to the conditions $p^2=0$ and $\a\slashed{p}=0$, where 
\be
\slashed{p}=p_\m\G^\m,
\ee
and our conventions for gamma matrices are detailed in appendix \ref{sec:GammaMatrixConventions}.  In this case there are gauge transformations parametrized by $\lambda_A$ and $\m_A$, generated by the operator
\be
U_{-\frac{3}{2}}=\ls -2i\lambda_Ac\p c\p\xi\p^2\xi e^{-\frac{7}{2}\phi}+2\sqrt{2}\m_Ac\p\xi e^{-\frac{5}{2}\phi}\rs S^Ae^{ipX},
\ee
which act as
\be
\d\a=\sqrt{2}\lambda\slashed{p},\qquad\d\beta=\lambda+\m\slashed{p}.
\ee
If we were to work in the relative cohomology $H_R^1$ instead, then we would set $\alpha=\lambda=0$ above, leading to an isomorphic cohomology at nonzero momentum, but with a smaller space of states and of gauge transformations.

Finally for picture $-\hlf$, we have
\be
V_{-\hlf}=\a_Ace^{-\hlf\phi}S^Ae^{ipX},
\ee
with $p^2=0$ and $\a\slashed{p}=0$.  There are no gauge transformations in this case, and the absolute and relative cohomologies are identical.

\subsubsection{Closed string}

The closed string vertex operators will, of course, be (sums of) products of left- and right-moving open string vertex operators.  In the NS-NS sector, we would have
\begin{equation}
V_{P, \tilde P} (z, \bar z)= \epsilon_{\mu \nu}\tilde V^\mu_P(z) V^{\nu} _{\tilde P} (\bar z)e^{i p X(z,\bar z)},
\end{equation}
where
\begin{equation}
\begin{split}
V^\mu_{-1} & = ce^{-\phi}\psi^\m,\\
V^\mu_0 & = c\lp\p X^\m-ip_\rho\psi^\rho\psi^\m\rp-\hlf e^\phi\eta\psi^\m,\\
\end{split}
\end{equation}
and similar expressions for the $\bar z$ dependent contributions. In each case BRST closure requires
\begin{equation}
\begin{split}
& p^2=0,\\
& p^\n\e_{\n\m}=p^\n\e_{\m\n}=0.\\
\end{split}
\end{equation}
There are gauge transformations
\be
\d\e_{\m\n}=\lambda_\m p_\n+p_\m\zeta_\n,
\ee
for vectors $\lambda_\m$ and $\zeta_\n$ satisfying $p^\m\lambda_\m=p^\m\zeta_\m=0$.  The operators constructed in this way satisfy $[b_0,V_{P,\tilde{P}}]=[\wtb_0,V_{P,\tilde{P}}]=0$, so we can consider them as elements of $\mathcal{H}_R^2$, $\mathcal{H}_S^2$, or $\mathcal{H}^2$.  Working in any of these three cohomologies we could enlarge the space of operators we consider to ones where the left and right ghost numbers were not both one (but the total ghost number should still be two), but by the arguments of section \ref{subsec:PhysicalStates} all such operators would then be BRST-trivial as long as $p^\m\ne 0$, so the ones we have written down here completely capture the cohomology at ghost number two.

In the R-R sector, we have the $(-\hlf,-\hlf)$-picture operator
\be
\label{eq:SymmetricRR}
V_{-\hlf,-\hlf}(z,\barz)=f_{AB}V_{-\hlf}^A(z)  \tilde V_{-\hlf}^B(\bar z) e^{ipX},
\end{equation}
where
\begin{equation}
V_{-\hlf}^A = ce^{-\hlf\phi}S^A
\end{equation}
In this case BRST implies
\begin{equation} \label{eq:RRBRSTcond}
\begin{split}
& p^2=0\\
&\slashed{p}^Tf=f\slashed{p}=0.
\end{split}
\end{equation}

We must take a moment now to discuss the GSO projection.  In the NS-NS sector, this is simply the requirement that $(-1)^F$ and $(-1)^{\widetilde{F}}$, where $F$ and $\widetilde{F}$ are the left and right world-sheet fermion number operators, are both equal to one when acting on a physical state.  It turns out that the NS-NS sector operators we wrote down already satisfy this requirement.  In the R-R sector, the GSO projection (we use the conventions of~\cite{Polchinski:1998rr}) is $(-1)^F=1$ and $(-1)^{\widetilde{F}}=(-1)^{p+1}$, where $p$ is even for IIA and odd for IIB.  The action of $(-1)^F$ on R sector ground states is given by
\bea
\label{eq:RamondStateFermionNumber}
\lp -1\rp^F\left|P;A\right\rangle &=& \lp -1\rp^{P+\hlf}\lp\G_{11}\rp^A_{\hphantom{A}B}\left|P;B\right\rangle,\\
\left\langle P;A\right|\lp -1\rp^F &=& \lp -1\rp^{P+\hlf}\lp\G_{11}\rp^A_{\hphantom{A}B}\left\langle P;B\right|.\non
\eea
where $P\in\zet+\hlf$ is the picture and $A$ is the spinor index.  The action of $(-1)^{\widetilde{F}}$ is given by the corresponding expressions with tildes.  Thus the GSO projection on (\ref{eq:SymmetricRR}) imposes
\be
f=\lp\G_{11}\rp^Tf=\lp -1\rp^{p+1}f\G_{11}.
\ee

These conditions determine the choice of coefficient $f_{AB}$.  Indeed,
there is a natural correspondence, using the algebra of gamma matrices ${(\Gamma^\mu)^A}_B$, between objects with two Lorentz spinor indices and (formal sums of) space-time differential forms according to
\be
f_{AB}=\lp\mathcal{C}\sum_n\frac{1}{n!}F^{(n)}_{\m_1\cdots\m_n}\G^{\m_1\cdots\m_n}\rp_{AB}=\lp\mathcal{C}\slashed{F}\rp_{AB},
\ee
We will now argue that the BRST and GSO conditions make it very natural to associate the differential forms $F^{(n)}$ with the R-R field strengths.  First of all, the GSO projection immediately gives that $F^{(n)}=0$ unless $p+n$ is even, so we only have even forms in IIA or odd forms in IIB.  The other implication of the GSO condition is the duality relation
\be
\ast F^{(n)}=\lp -1\rp^{\hlf(n^2-n)}F^{(10-n)},
\ee
which comes from $\slashed{F}=-\G_{11}\slashed{F}$.

Turning next to the BRST conditions, it is easy to check that $\slashed{p}^Tf=f\slashed{p}=0$ is equivalent to the demand that
\be
dF^{(n)}=d\ast F^{(n)}=0,
\ee
for each $n$.  Finally, since there were no BRST gauge transformations, the $F^{(n)}$ should be gauge invariant quantities.  It is thus very natural to associate these $F^{(n)}$ with the R-R field strengths, at least up to an overall normalization which we won't attempt to fix.


For picture $(-\frac{3}{2},-\hlf)$, if we work in the absolute cohomology $\mathcal{H}^2$, we have~\cite{Callan:1986bc} (see also discussions of R-R vertex operators in~\cite{Billo:1998vr,Liu:2001qa})
\be
V_{-\frac{3}{2},-\hlf}(z,\barz)=\ls f_{AB}\left( c\p c\p\xi e^{-\frac{5}{2}\phi}S^A\right) \left( \wtc e^{-\hlf\wtphi}\widetilde{S}^B\right)+ig_{AB}\left( ce^{-\frac{3}{2}\phi}S^A\right) \left( \wtc e^{-\hlf\wtphi}\widetilde{S}^B\right)\rs e^{ipX},
\ee
with
\begin{equation}
\begin{split}
& p^2 =0,\\
& \slashed{p}^Tf  =f\slashed{p}=0,\\
& g\slashed{p} =0,\\
\end{split}
\end{equation}
and where we have gauge transformations
\be
\begin{split} \label{bi}
& \d f=\sqrt{2}\slashed{p}^T\zeta,\\
& \d g=\zeta+\slashed{p}^T\chi,\\
\end{split}
\ee
for any parameters $\zeta_{AB}$ and $\chi_{AB}$ satisfying $\zeta\slashed{p}=\chi\slashed{p}=0$.  As before, $f_{AB}$ corresponds to a gauge-invariant differential form which is closed and co-closed, so it should be proportional to $\mathcal{C}\slashed{F}$.  

For the other term, writing $g=\mathcal{C}\slashed{G}$ for a sum of differential forms $G^{(n)}$, the GSO projection becomes
\be
\slashed{G}=\G_{11}\slashed{G}=\lp -1\rp^p\slashed{G}\G_{11},
\ee
which implies that we must only have terms satisfying $p+n$ is odd, and we must have
\be
\ast G^{(n)}=\lp -1\rp^{\hlf(n^2-n+2)}G^{(10-n)}.
\ee
The BRST condition $g\slashed{p}=0$ becomes the constraint
\be
dG^{(n)}=-\ast d\ast G^{(n+2)}.
\ee

The gauge transformations imply that we can make shifts (these are the gauge transformations parametrized by $\chi$)
\be
\label{eq:GGaugeTransformation}
\d G^{(n)}=d\Lambda^{(n-1)},
\ee
where the forms $\Lambda^{(n-1)}$ satisfy
\be
d\Lambda^{(n-1)}=-\ast d\ast\Lambda^{(n+1)}.
\ee
By using the gauge transformation $\zeta_{AB}=-g_{AB}$ we see that we can replace $g_{AB}$ by a contribution to $f_{AB}$ given by $\d F^{(n)}=2\sqrt{2}dG^{(n-1)}$.  For this reason it is natural to assume that $G^{(n)}$ should b proportional to the R-R potential $C^{(n)}$ in a specific gauge.  Indeed, we should write
\be
\label{eq:RRcoeffs}
f_{AB}=2\lp 1-y\rp\lp\mathcal{C}\slashed{F}\rp_{AB},\qquad g_{AB}=\frac{y}{\sqrt{2}}\lp\mathcal{C}\slashed{C}\rp_{AB}.
\ee
Here $y$ is an arbitrary real parameter, and we have chosen a gauge in which $dC^{(n-1)}=-\ast d\ast C^{(n+1)}$, i.e.
\be
\label{eq:RRGaugeChoice}
C^{(n)}_{\m_1\cdots\m_{n-1}\n}p^\n=-(n-1)C^{(n-2)}_{[\m_1\cdots\m_{n-2}}p_{\m_{n-1}]}.
\ee
The overall factor of two in (\ref{eq:RRcoeffs}) ensures that amplitudes using this operator agree with those computed using (\ref{eq:SymmetricRR}).


Our gauge transformations then act to either shift the parameter $y$ by any amount, or to shift $\slashed{C}$ according to (\ref{eq:GGaugeTransformation}).
The most convenient choice of gauge for us is to simply take $y$=0, in which case we can simply write
\be
\label{eq:y=0Gauge}
V_{-\frac{3}{2},-\hlf}(z,\barz)=f_{AB}V_{-\frac{3}{2} }^A \tilde V_{-\hlf} ^B e^{i p X} ,
\ee
where
\begin{equation}
V_{-\frac{3}{2} }^A=c\p c\p\xi e^{-\frac{5}{2}\phi}S^A, \qquad \tilde V_{-\hlf} ^B= \wtc e^{-\hlf\wtphi}\widetilde{S}^B.
\end{equation}
Of course, if we want to use our vertex operators on arbitrary Riemann surfaces, then we should work with $\mathcal{H}_S^2$ rather than $\mathcal{H}^2$, and so we should be forced to take $y=1$ and disallow the gauge transformations which shift $y$.  However, we will demonstrate in section \ref{sec:BRSTExact} that for two-point disc amplitudes, we encounter no problems working with the more general form, and it is somewhat more convenient for calculation (we don't need to deal with the messy gauge condition (\ref{eq:RRGaugeChoice}) for instance).


The story for picture $(-\hlf,-\frac{3}{2})$ is completely analogous.


\section{Boundary states and correlators}
\label{sec:BoundaryState}

We will try to develop all the properties and results for our boundary states from scratch, but some other useful references include~\cite{Callan:1986bc,Kostelecky:1987px} and the review~\cite{DiVecchia:1999rh}.

\subsection{Boundary states}

We will be taking a very pragmatic view of the boundary state $|B\rangle$ as being simply an implementation of the boundary conditions obeyed by the vertex operators.  These boundary conditions relate right-moving excitations to left-moving excitations, as waves hit the boundary of the string world-sheet and reflect back.  On the upper half-plane, we expect the boundary conditions to relate a purely right-moving operator of conformal weight $h$ to a purely left-moving counterpart,
\be
\widetilde{\mathcal{O}}^a(\bar{w})=\eta^{2h}R^a_b\mathcal{O}^b(w)|_{w=\bar{w}}.
\ee
Here $a$ and $b$ are indices that correspond to the Lorentz representation of the operator, and $R^a_b$ is a matrix encoding the boundary conditions.  The sign $\eta=\pm$ is included for operators of half-integral conformal weight so that we can later sum over the two choices when performing the GSO projection.

We will find it more useful to work not on the upper half-plane, but on the exterior of the unit disc, $|z|>1$ (this is so that we can place the boundary state at the origin and propagate it outwards).  In this case, if $\widetilde{\mathcal{O}}$ is a primary conformal operator, then under the mapping $z=e^{-iw}$, the condition above becomes
\be
\lp i\barz\rp^h\widetilde{\mathcal{O}}^a(\barz)=R^a_b\eta^{2h}\lp -iz\rp^h\mathcal{O}^b(z)|_{|z|^2=1}.
\ee
The boundary state is designed to relate an anti-holomorphic operator defined on the exterior of the disc to a holomorphic operator defined on the interior of the disc in a way consistent with the boundary conditions,
\be
\label{eq:MasterEq}
\widetilde{\mathcal{O}}^a(\barz)\left|B;\eta\right\rangle=R^a_b\lp i\eta\barz\rp^{-2h}\mathcal{O}^b(\barz^{-1})\left|B;\eta\right\rangle.
\ee


In the case that the operators in (\ref{eq:MasterEq}) correspond to free fields,
\be
\mathcal{O}^a(z)=\Phi^a(z)=\sum_{r\in\zet-h}\Phi^a_rz^{-r-h},\qquad\widetilde{\mathcal{O}}^a(\barz)=\widetilde{\Phi}^a(\barz)=\sum_{r\in\zet-h}\widetilde{\Phi}^a_r\barz^{-r-h},
\ee
then (\ref{eq:MasterEq}) implies boundary conditions on oscillators given by
\be
\label{eq:FreeFieldMaster}
\widetilde{\Phi}^a_r\left|B;\eta\right\rangle=\lp i\eta\rp^{-2h}R^a_b\Phi^b_{-r}\left|B;\eta\right\rangle.
\ee

Note that these expressions are valid only for {\it{primary}} conformal fields, and need to be modified otherwise.  For instance, the current which measures $\phi$ charge, $J_\phi=-\p\phi$, obeys
\be
T(z)J_\phi(0)\sim\frac{2}{z^3}+\frac{J_\phi(0)}{z^2}+\frac{\p J_\phi(0)}{z}.
\ee
Due to the $z^{-3}$ term above, the transformation of $J_\phi$ under the mapping $z=e^{-iw}$ is given by
\be
J_\phi(w)=-izJ_\phi(z)-i,
\ee
and so the action on the boundary state should be given by
\be
\widetilde{J}_{\wtphi}(\barz)\left|B;\eta\right\rangle=\lp -\barz^{-2}J_\phi(\barz^{-1})-2\barz^{-1}\rp\left|B;\eta\right\rangle.
\ee
From this it follows that the total left- plus right- $\phi$ charge of the boundary state must be $-2$,
\be
\label{eq:BPhiCharge}
\lp\oint\frac{dz}{2\pi i}J_\phi(z)-\oint\frac{d\barz}{2\pi i}\widetilde{J}_{\wtphi}(\barz)\rp\left|B;\eta\right\rangle=-2\left|B;\eta\right\rangle.
\ee

The property (\ref{eq:MasterEq}) essentially fixes the boundary state $|B;\eta\rangle$ up to multiplication by an overall, possibly $\eta$-dependent, number.  We will try to avoid using the explicit form of $|B;\eta\rangle$ whenever possible, simply making repeated use of (\ref{eq:MasterEq}).  

We mention here one more aspect which will be of use in evaluating the correlators below.  Let us schematically write the boundary state as
\be
\left|B\right\rangle=\sum_{\vec{n},\vec{m}\ge 0}B_{\vec{n}\vec{m}}\left|\vec{n},\vec{m}\right\rangle,
\ee
where $\vec{n}$ and $\vec{m}$ label states on the left and right respectively.  We will use the label $0$ for the vacuum states, and we consider an ordering where $\vec{n}_1>\vec{n}_2$ if $|\vec{n}_1\rangle$ is obtained from $|\vec{n}_2\rangle$ by acting with creation operators.  Then, again schematically, if we act with a left-moving annihilation operator $\alpha_{\vec{r}}$ labeled by $\vec{r}>0$ we get
\be
\alpha_{\vec{r}}\left|B\right\rangle=\sum_{\vec{m}\ge 0}\sum_{\vec{n}\ge\vec{r}}B_{\vec{n}\vec{m}}\left|\vec{n}-\vec{r},\vec{m}\right\rangle.
\ee
By property (\ref{eq:FreeFieldMaster}), we should be able to equivalently write this in terms of a right-moving creation operator, $\widetilde{\alpha}_{-\vec{r}}$,
\be
\widetilde{\alpha}_{-\vec{r}}\left|B\right\rangle=\sum_{\vec{n},\vec{m}\ge 0}B_{\vec{n}\vec{m}}\left|\vec{n},\vec{m}+\vec{r}\right\rangle.
\ee
Comparing these two expressions for arbitrary $\vec{r}>0$, we see that the second series does not contain the state $|0,0\rangle$, and so in the first series we must have $B_{\vec{r}0}=0$.  We can similarly derive that $B_{0\vec{r}}=0$.  The key implication of this result is that
\be
\label{eq:BVacuumReduction}
\left\langle\vec{n},0|B\right\rangle=B_{00}\left\langle\vec{n}|0\right\rangle,
\ee
where $B_{00}$ is just a number, or, if the vacua are degenerate, a matrix.  In this case, if we label vacua by an extra index $\alpha$, then we have
\be
\label{eq:BDegenerateVacua}
\left\langle\vec{n},0_\alpha|B\right\rangle=\sum_\beta \lp B_{00}\rp_{\beta\alpha}\left\langle\vec{n}|0_\beta\right\rangle.
\ee

This property means that in a correlator, if we first use (\ref{eq:MasterEq}) to rewrite all the operators as holomorphic, we can then simply evaluate a holomorphic correlator with a vacuum as in-state, and do not need to worry about the detailed structure of the boundary state, except the zero-mode part which accounts for degenerate vacua.

\subsection{Correlators}
\label{subsec:Correlators}

In this subsection we will use (\ref{eq:MasterEq}) to evaluate various correlators that we will need to compute disc amplitudes, proceeding sector by sector.  When we eventually compute full amplitudes, we will find it convenient to simplify our calculations by sending the position of the first vertex operator, $z_1$, to infinity (the full amplitude is independent of $z_1$).  In this limit, it is natural to absorb any spin fields and exponentials of $\phi$ or $\wtphi$ into the out-state using
\be
\lim_{z\rightarrow\infty}\vphantom{\left\langle 0_\psi;(0,0)_\phi\right|}_{NS}\left\langle 0_\psi;(0,0)_\phi\right|:e^{Q\phi(z)}::e^{\widetilde{Q}\wtphi(\barz)}:\simeq z^{Q(Q+2)}\barz^{\widetilde{Q}(\widetilde{Q}+2)}\vphantom{\left\langle 0_\psi;(Q,\widetilde{Q})_\phi\right|}_{NS}\left\langle 0_\psi;(Q,\widetilde{Q})_\phi\right|,
\ee
for integral $Q$ and $\widetilde{Q}$, and
\begin{multline}
\lim_{z\rightarrow\infty}\vphantom{\left\langle 0_\psi;(0,0)_\phi\right|}_{NS}\left\langle 0_\psi;(0,0)_\phi\right| :e^{Q\phi(z)}S^A(z)::e^{\widetilde{Q}\wtphi(\barz)}\widetilde{S}^B(\barz): \\
\simeq z^{Q(Q+2)-\frac{5}{4}}\barz^{\widetilde{Q}(\widetilde{Q}+2)-\frac{5}{4}}\vphantom{\left\langle (A,B)_\psi;(Q,\widetilde{Q})_\phi\right|}_R\left\langle (A,B)_\psi;(Q,\widetilde{Q})_\phi\right|,
\end{multline}
for half-integral $Q$ and $\widetilde{Q}$.

We will now discuss some of the particular correlators that we will need in different sectors.

\subsubsection{$X$ sector}

To relate the antiholomorphic fields to the holomorphic fields we use (\ref{eq:MasterEq}) with $R^\m_\n=D^\m_{\hphantom{\m}\n}$, the diagonal matrix with entries $+1$ for directions along the brane and $-1$ for orthogonal directions.  If we use indices $a,b,\cdots$ and $i,j,\cdots$ for these tangent and normal directions respectively, then (lowering the index with $\eta_{\m\n}$), we have $D_{ab}=\eta_{ab}$, $D_{ai}=D_{ia}=0$, $D_{ij}=-\d_{ij}$.  Then we find for example,
\be
\overline{\p}X^\m(\barz)\left|B\right\rangle=-\barz^{-2}D^\m_{\hphantom{\m}\n}\p X^\n(\barz^{-1})\left|B\right\rangle.
\ee

For exponentials we first split into holomorphic and anti-holomorphic parts,
\be
e^{ipX(z,\barz)}=e^{ipX(z)}e^{ip\widetilde{X}(\barz)},
\ee
and then we use the boundary state to convert the anti-holomorphic piece into a holomorphic operator,
\be
e^{ip\widetilde{X}(\barz)}\left|B\right\rangle=e^{ipDX(\barz^{-1})}\left|B\right\rangle,
\ee
where we assume here $p^2=0$, so that the exponential has zero conformal weight.  

Once we have converted all the antiholomorphic operators in a correlator into holomorphic ones using (\ref{eq:MasterEq}), then we can use (\ref{eq:BVacuumReduction}) to evaluate the correlator.  In this sector there are degenerate vacua, labeled by momenta, but as usual for non-compact directions we must have left and right momenta equal\footnote{Thus the zero-mode part of this boundary state has
\be
B_{00}(k,k')=B_{00}(k)\d^{10}(k-k').
\ee
Furthermore, applying (\ref{eq:FreeFieldMaster}) to the momentum operator we learn that
\be
k_\m B_{00}(k)=-\lp Dk\rp_\m B_{00}(k),
\ee
which implies that $B_{00}(k)$ is nonzero only for momenta transverse to the brane.  We won't actually need this result however, since we can use (\ref{eq:MasterEq}) to reduce everything to the case with $k=0$.}, and since we take the out-state to have zero momentum ($k_\alpha=0$ in the notation of (\ref{eq:BDegenerateVacua})), the correlator will simply reduce to a holomorphic correlator.

It is now straightforward to evaluate the expectation value for products of exponentials,
\begin{multline}
\left\langle 0_X\left|:e^{ip_1X(z_1,\barz_1)}:\cdots:e^{ip_nX(z_n,\barz_n)}:\right|B_X\right\rangle=\lp 2\pi\rp^{p+1}\d^{p+1}(\hlf\lp 1+D\rp\sum_{i=1}^np_i)\\
\times\prod_{k=1}^n\lp\left|z_k\right|^2-1\rp^{p_kDp_k}
\prod_{1\le\ell<m\le n}\left|z_\ell-z_m\right|^{2p_\ell p_m}\left|z_\ell\barz_m-1\right|^{2p_\ell Dp_m}.
\end{multline}
Indeed the second line appears frequently enough that we shall abbreviate it with the symbol $\mathcal{K}$.  The first line implements conservation of momentum along the brane; in the transverse directions we do not have conservation of momentum\footnote{Alternatively, we can think of the D-brane itself, or equivalently the boundary state, as carrying momentum in the transverse directions, as in the previous footnote.}.

We will also need correlators which include explicit factors of $\p X^\m$ or $\bar{\p}X^\m$.  Again, we convert right-movers into left-movers using the boundary state, and then evaluate the correlator using the usual methods.  For example,
\begin{multline}
\left\langle 0_X\left|:e^{ip_1X(z_1,\barz_1)}:\cdots :e^{ip_{n-1}X(z_{n-1},\barz_{n-1})}::\p X^\m(z_n)e^{ip_nX(z_n,\barz_n)}:\right|B_X\right\rangle\\
=\left\langle 0_X\left|:e^{ip_1X(z_1,\barz_1)}:\cdots :e^{ip_nX(z_n,\barz_n)}:\right|B_X\right\rangle\\
\times\lp\frac{ip_1}{z_1-z_n}+\cdots +\frac{ip_{n-1}}{z_{n-1}-z_n}-\frac{i\barz_1Dp_1}{z_n\barz_1-1}-\cdots -\frac{i\barz_nDp_n}{\left|z_n\right|^2-1}\rp^\m.
\end{multline}

\subsubsection{$bc$ sector}

Applying (\ref{eq:MasterEq}) here gives
\be
\wtc(\barz)\left|B\right\rangle=-\barz^2c(\barz^{-1})\left|B\right\rangle,\qquad\wtb(\barz)\left|B\right\rangle=\barz^{-4}b(\barz^{-1})\left|B\right\rangle,
\ee
or simply
\be
\wtc_n\left|B\right\rangle=-c_{-n}\left|B\right\rangle,\qquad\wtb_n\left|B\right\rangle=b_{-n}\left|B\right\rangle.
\ee
The boundary state which implements these relations is given by\footnote{Again, as throughout this work we are not fixing the overall normalization of the amplitudes, including the normalization of the boundary state.}
\be
\left| B_{bc}\right\rangle=\exp\ls\sum_{n=1}^\infty\lp c_{-n}\wtb_{-n}-b_{-n}\wtc_{-n}\rp\rs\frac{c_0+\wtc_0}{2}c_1\wtc_1\left|0_{bc}\right\rangle.
\ee

Because of the insertion of $(b_0+\wtb_0)$ in front of the boundary state, it is actually simpler to just list the  correlator that we will need,
\be
\left\langle 0_{bc}\left|c(z_1)\wtc(\barz_1)c(z_2)\wtc(\barz_2)\lp b_0+\wtb_0\rp\right|B_{bc}\right\rangle=\left|z_1-z_2\right|^2\lp\left|z_1z_2\right|^2-1\rp,
\ee
\be
\left\langle 0_{bc}\left|c\p c(z_1)\wtc(\barz_1)c(z_2)\lp b_0+\wtb_0\rp\right|B_{bc}\right\rangle=\barz_1\lp z_1-z_2\rp^2,
\ee
\be
\left\langle 0_{bc}\left|c\p c(z_1)\wtc(\barz_1)\wtc(\barz_2)\lp b_0+\wtb_0\rp\right|B_{bc}\right\rangle=\lp\barz_1-\barz_2\rp\lp z_1^2\barz_1\barz_2-1\rp.
\ee

\subsubsection{$\phi$ sector}

In this sector we will evaluate correlators which are products of exponentials of $\phi$ or $\wtphi$.  Since $e^{\widetilde{Q}\wtphi}$ is a primary operator of dimension $-\hlf\widetilde{Q}(\widetilde{Q}+2)$, we can use our techniques above to convert it to an exponential of $\phi$.  Indeed, (\ref{eq:MasterEq}) with $R^\phi_\phi=1$ will yield
\be
e^{\widetilde{Q}\wtphi(\barz)}\left|B;\eta\right\rangle=\lp i\eta\barz\rp^{\widetilde{Q}\lp\widetilde{Q}+2\rp}e^{\widetilde{Q}\phi(\barz^{-1})}\left|B;\eta\right\rangle.
\ee

In this sector there are degenerate vacua labeled by the the picture $P$.  From (\ref{eq:BPhiCharge}), the only non-zero matrix elements of $B_{00}$ will be those correspond to total picture charge $-2$.  After converting all of the anti-holomorphic exponentials into holomorphic ones, we will pick out only the piece with picture $(\widetilde{Q}_1,-2-\widetilde{Q}_1)$.  To make use of (\ref{eq:BDegenerateVacua}) we still need the constants $B_{00}(\widetilde{Q}_1,-2-\widetilde{Q}_1)$.  There is an overall constant which we cannot determine, but by imposing consistency between pictures (the freedom to have rewritten $e^{\widetilde{Q}_1\wtphi(\barz_1)}$ as a holomorphic insertion, one can show that 
\be
B_{00}(\widetilde{Q}_1,-2-\widetilde{Q}_1)\sim\lp i\eta\rp^{(\widetilde{Q}_1-a)(\widetilde{Q}_1-a+2)},
\ee
where $a$ is a constant that should be an integer in the NS sector or half-integer in the R sector.  We have a choice of what value of $a$ to take in each sector; we will choose $a=-1$ and $a=-1/2$ in the NS and R sector respectively.  

Taking all of this into account, we can then derive the correlator of an arbitrary number of exponentials of $\phi$ and $\wtphi$,
\begin{multline}
\label{eq:GeneralPhiResult}
\left\langle Q_1,\widetilde{Q}_1\left|:e^{Q_2\phi(z_2)}::e^{\widetilde{Q}_2\widetilde{\phi}(\bar{z}_2)}:\cdots :e^{Q_n\phi(z_n)}::e^{\widetilde{Q}_n\widetilde{\phi}(\bar{z}_n)}:\right|B_\phi;\eta\right\rangle= \\
\lp i\eta\rp^{(\widetilde{Q}_1-a)(\widetilde{Q}_1-a+2)+\sum_{k=2}^n\widetilde{Q}_k(\widetilde{Q}_k+2)}e^{i\pi(\widetilde{Q}_1-a)(2+Q_1+\widetilde{Q}_1)}\lp\prod_{k=2}^nz_k^{-Q_k\widetilde{Q}_1}\bar{z}_k^{-\widetilde{Q}_kQ_1}\lp\left|z_k\right|^2-1\rp^{-Q_k\widetilde{Q}_k}\rp \\
\times\lp\prod_{1<i<j\le n}\lp z_i-z_j\rp^{-Q_iQ_j}\lp\bar{z}_j-\bar{z}_i\rp^{-\widetilde{Q}_i\widetilde{Q}_j}\lp z_i\bar{z}_j-1\rp^{-Q_i\widetilde{Q}_j}\lp 1-\bar{z}_iz_j\rp^{-\widetilde{Q}_iQ_j}\rp,
\end{multline}


\subsubsection{$\eta\xi$ sector}

This sector will be dealt with on a case by case basis, and we will make use of (\ref{eq:MasterEq}) with $R=1$, so that
\be
\widetilde{\eta}(\barz)\left|B\right\rangle=-\barz^{-2}\eta(\barz^{-1})\left|B\right\rangle,\qquad\widetilde{\xi}(\barz)\left|B\right\rangle=\xi(\barz^{-1})\left|B\right\rangle.
\ee

\subsubsection{$\psi$ sector}

For NS sector amplitudes, there are not degenerate vacua, so up to an undetermined normalization we will simply use our rule to convert $\wtpsi^\m$ into $\psi^\m$ using (\ref{eq:MasterEq}),
\be
\label{eq:PsiMaster}
\wtpsi^\m(\barz)\left|B;\eta\right\rangle_{NS}=-i\eta\barz^{-1}\lp D\psi\rp^\m(\barz^{-1})\left|B;\eta\right\rangle_{NS},
\ee
and then use the OPE (\ref{eq:PlaneCorrelators}), so for example
\be
{\vphantom{\left\langle 0_\psi\left|\psi^\m(z_1)\wtpsi^\n(\barz_2)\right|B_\psi;\eta\right\rangle}}_{NS}\left\langle 0_\psi\left|\psi^\m(z_1)\wtpsi^\n(\barz_2)\right|B_\psi;\eta\right\rangle_{NS}=\frac{-i\eta D^{\m\n}}{z_1\barz_2-1}.
\ee

In the R sector, there are zero modes $\psi^\m_0$, and by (\ref{eq:FreeFieldMaster}) these should obey
\be
\label{eq:PsiZeroModeBC}
\wtpsi^\m_0\left|B;\eta\right\rangle_R=-i\eta\lp D\psi\rp^\m_0\left|B;\eta\right\rangle_R.
\ee
These zero modes lead to degenerate vacua labeled by spinor indices $A$, $B$, etc (see our spinor and gamma matrix conventions in appendix \ref{sec:GammaMatrixConventions}).  Let us write
\be
\mathcal{M}(\eta)_{AB}=\lp B_{00}\rp_{AB}.
\ee
Since the zero modes of $\psi$ and $\wtpsi$ act on R ground states as
\be
\psi^\m_0\left|A,B\right\rangle=\frac{1}{\sqrt{2}}\lp\G^\m\rp^A_{\hphantom{A}C}\left|C,B\right\rangle,\qquad\wtpsi^\m_0\left|A,B\right\rangle=\frac{1}{\sqrt{2}}\lp\G_{11}\rp^A_{\hphantom{A}C}\lp\G^\m\rp^B_{\hphantom{B}D}\left|C,D\right\rangle,
\ee
we can rewrite (\ref{eq:PsiZeroModeBC}) as
\be
\label{eq:MRelation}
\lp\G_{11}\rp^T\mathcal{M}(\eta)\G^\m=-i\eta D^\m_{\hphantom{\m}\n}\lp\G^\n\rp^T\mathcal{M}(\eta).
\ee
It is not difficult to check that this relation is solved by
\be
\label{eq:MExpression}
\mathcal{M}(\eta)=\lp i\eta\rp^p\mathcal{C}\G^0\cdots\G^p\lp P_+-i\eta P_-\rp,
\ee
with
\be
P_\pm=\hlf\lp 1\pm\G_{11}\rp,
\ee
$\mathcal{C}_{AB}$ is an antisymmetric charge-conjugation matrix, and we have assumed that the D$p$-brane is extended in the directions $0$ through $p$.  The boundary conditions only fix $\mathcal{M}(\eta)$ up to an overall, possibly $\eta$-dependent constant.  We have chosen the prefactor $(i\eta)^p$ for later convenience.

This, along with the holomorphic expectation values for products of $\psi$ between R ground states, leads to the result
\begin{multline}
\label{eq:ProductOfPsi}
{\vphantom{\left\langle A,B\left|\psi^{\m_1}(z_1)\cdots\psi^{\m_n}(z_n)\right|B_\psi;\eta\right\rangle}}_R\left\langle A,B\left|\psi^{\m_1}(z_1)\cdots\psi^{\m_n}(z_n)\right|B_\psi;\eta\right\rangle_R=\lp -1\rp^{n+1}2^{-\frac{n}{2}}\lp z_1\cdots z_n\rp^{-\hlf} \\
\times\left\{\ls\G^{\m_1\cdots\m_n}\mathcal{C}^{-1}\mathcal{M}(\eta)\mathcal{C}^{-1}\rs^{AB}+\frac{z_1+z_2}{z_1-z_2}\eta^{\m_1\m_2}\ls\G^{\m_3\cdots\m_n}\mathcal{C}^{-1}\mathcal{M}(\eta)\mathcal{C}^{-1}\rs^{AB}\right. \\
\left.+\cdots+\frac{z_1+z_2}{z_1-z_2}\frac{z_3+z_4}{z_3-z_4}\eta^{\m_1\m_2}\eta^{\m_3\m_4}\ls\G^{\m_5\cdots\m_n}\mathcal{C}^{-1}\mathcal{M}(\eta)\mathcal{C}^{-1}\rs^{AB}+\cdots\right\},
\end{multline}
where $\cdots$ represent all other possible contractions, with appropriate signs from anticommuting the fermions or the gamma matrices.  Then any desired correlator can be obtained by first using (\ref{eq:PsiMaster}) and then using (\ref{eq:ProductOfPsi}).

As examples we have
\be
{\vphantom{\left\langle A,B\left|\psi^\m(z)\right|B_\psi;\eta\right\rangle}}_R\left\langle(A,B)_\psi\left|\psi^\m(z)\right|B_\psi;\eta\right\rangle_R=\frac{1}{\sqrt{2z}}\ls\G^\m\mathcal{C}^{-1}\mathcal{M}(\eta)\mathcal{C}^{-1}\rs^{AB},
\ee
and
\begin{multline}
{\vphantom{\left\langle A,B\left|\psi^\m(z)\wtpsi^\n\wtpsi^\rho(\barz)\right|B_\psi;\eta\right\rangle}}_R\left\langle A,B\left|\psi^\m(z)\wtpsi^\n\wtpsi^\rho(\barz)\right|B_\psi;\eta\right\rangle_R=-\frac{1}{2\barz\sqrt{2z}}D^\n_{\hphantom{\n}\s}D^\rho_{\hphantom{\rho}\tau}\left\{\ls\G^{\m\s\tau}\mathcal{C}^{-1}\mathcal{M}(\eta)\mathcal{C}^{-1}\rs^{AB}\right.\\
\left.+\frac{|z|^2+1}{|z|^2-1}\lp\eta^{\m\s}\ls\G^\tau\mathcal{C}^{-1}\mathcal{M}(\eta)\mathcal{C}^{-1}\rs^{AB}-\eta^{\m\tau}\ls\G^\s\mathcal{C}^{-1}\mathcal{M}(\eta)\mathcal{C}^{-1}\rs^{AB}\rp\right\}.
\end{multline}

\subsection{Evaluating the Traces}
\label{subsec:Traces}

From the $\psi$ correlators in the R-R sector above, we will find our amplitudes include traces of the form
\be
T^{\m_1\cdots\m_n}=f_{AB}\ls\G^{\m_1\cdots\m_n}\mathcal{C}^{-1}\mathcal{M}(\eta)\mathcal{C}^{-1}\rs^{AB}=\Tr\ls f\mathcal{C}^{-1}\mathcal{M}(\eta)^T\mathcal{C}^{-1}\lp\G^{[\m_n}\rp^T\cdots\lp\G^{\m_1]}\rp^T\rs
\ee
Using the explicit form of $\mathcal{M}(\eta)$ from (\ref{eq:MExpression}), the relation (\ref{eq:ChargeTranspose}), and writing $f=\mathcal{C}\slashed{F}$, this becomes
\be
T^{\m_1\cdots\m_n}=\lp i\eta\rp^p\lp -1\rp^{\hlf(p^2-p+n^2+n)}\Tr\ls\slashed{F}\lp P_--i\eta P_+\rp\G^{0\cdots p}\G^{\m_1\cdots\m_n}\rs.
\ee
If we now bring in the fact that the GSO projection on R-R field strengths implies $\slashed{F}\G_{11}=(-1)^{p+1}\slashed{F}$, we find (for instance by separately evaluating for $p$ even and $p$ odd)
\be
T^{\m_1\cdots\m_n}=\lp -1\rp^{\hlf(n^2+n)}\Tr\ls\slashed{F}\G^{0\cdots p}\G^{\m_1\cdots\m_n}\rs.
\ee
Finally, if we use $a$ to denote indices along the brane and $i$ to denote transverse indices, then the trace picks out only the field strength of degree $p+1+\ell-k$ and we have explicitly
\be
\label{eq:ExplicitTraceFormula}
T^{a_1\cdots a_ki_1\cdots i_\ell}=\lp -1\rp^{\hlf(p^2+p+k^2+k)+p\ell+1}\frac{32}{(p+1-k)!}\e^{a_1\cdots a_k}_{\hphantom{a_1\cdots a_k}b_1\cdots b_{p+1-k}}F^{b_1\cdots b_{p+1-k}i_1\cdots i_\ell}.
\ee

\subsection{GSO Projection}
\label{subsec:GSO}

Finally, we should apply the GSO projection to our boundary states before inserting them into amplitudes.  In the NS sector, this means taking
\be
\label{eq:NSGeneralGSO}
\left|B\right\rangle_{NS}=\frac{1+\lp -1\rp^F}{2}\frac{1+\lp -1\rp^{\widetilde{F}}}{2}\left|B;+\right\rangle_{NS},
\ee
where $F$ and $\widetilde{F}$ are the left-moving and right-moving world-sheet fermion numbers, and in the R sector we take
\be
\label{eq:RGeneralGSO}
\left|B\right\rangle_R=\frac{1+\lp -1\rp^F}{2}\frac{1+\lp -1\rp^{p+1}\lp -1\rp^{\widetilde{F}}}{2}\left|B;+\right\rangle_R.
\ee

In appendix \ref{sec:BoundaryStateFermionNumber} we show that in our conventions the fermion numbers act on the boundary states as
\be
\label{eq:NSBoundaryStateFermion}
\lp -1\rp^F\left|B;\eta\right\rangle_{NS}=-\left|B;-\eta\right\rangle_{NS},\qquad\lp -1\rp^{\widetilde{F}}\left|B;\eta\right\rangle_{NS}=-\left|B;-\eta\right\rangle_{NS},
\ee
and
\be
\label{eq:RBoundaryStateFermion}
\lp -1\rp^F\left|B;\eta\right\rangle_R=\left|B;-\eta\right\rangle_R,\qquad\lp -1\rp^{\widetilde{F}}\left|B;\eta\right\rangle_R=\lp -1\rp^{p+1}\left|B;-\eta\right\rangle_R.
\ee
These then imply that the correct GSO-projected boundary states are
\be
\left|B\right\rangle_{NS}=\hlf\lp\left|B;+\right\rangle_{NS}-\left|B;-\right\rangle_{NS}\rp,
\ee
and
\be
\left|B\right\rangle_R=\hlf\lp\left|B;+\right\rangle_R+\left|B;-\right\rangle_R\rp.
\ee

\subsection{Amplitudes}

Finally, inside amplitudes we must also insert a ghost factor $(b_0+\widetilde{b}_0)$ and a propagator which pushes the boundary out to the first insertion point, so the total state is given by
\be
\lp b_0+\widetilde{b}_0\rp\int_{|w|>\operatorname{max}\{1/|z_i|\}}\frac{d^2w}{|w|^2}w^{-L_0}\barw^{-\widetilde{L}_0}\left|B\right\rangle.
\ee
Here $z_i$ are the insertion points of the various operators.  Some of these may be integrated over (for three- or higher-point functions), in which case the $w$ integration as defined should be taken as the inner-most integral.


\section{BRST-Exact States}
\label{sec:BRSTExact}

We will now discuss certain features of amplitudes on the disc.  We will show that BRST-exact operators do not necessarily decouple from such amplitudes.  However, we will then demonstrate that if all the operators in the amplitude correspond to states with generic momenta, so that we can use analytic continuation of momenta to do the computation, then the BRST-exact states do decouple and the amplitude should be gauge invariant.  With this result we can also show that, under the same assumptions, the amplitude does not depend on how we distribute picture charge.  We will show these results for the two-point functions on the disc, since those are the ones relevant in the current paper, but most of these arguments will carry forward to higher-point functions~\cite{inprogress}.

To construct the two-point function, we insert a pair of BRST-closed operators at arbitrary fixed positions on the sphere, and also insert the appropriate boundary state along with a ghost factor $(b_0+\widetilde{b}_0)$ and a propagator which extends the boundary state out to the first operator insertion.  We can use the conformal group of the sphere to fix the position of the boundary state insertion at $z=0$ on the complex plane, and then the amplitude is written
\be
\left\langle\left. V^{(1)}(z_1,\bar{z}_1)V^{(2)}(z_2,\bar{z}_2)(b_0+\widetilde{b}_0)\int_{|w|>\operatorname{max}(1/|z_1|,1/|z_2|)}\frac{d^2w}{|w|^2}w^{-L_0}\bar{w}^{-\widetilde{L}_0}\right|B\right\rangle.
\ee
We can then pull the integration to the left and use the relation
\be
w^{L_0}\bar{w}^{\widetilde{L}_0}\mathcal{O}(z,\bar{z})w^{-L_0}\bar{w}^{-\widetilde{L}_0}=w^h\bar{w}^{\widetilde{h}}\mathcal{O}(zw,\bar{z}\bar{w}),
\ee
for an operator $\mathcal{O}$ of conformal weight $(h,\widetilde{h})$, to express the amplitude as
\be
\label{eq:TwoPoint}
\int_{|w|>\operatorname{max}(1/|z_1|,1/|z_2|)}\frac{d^2w}{|w|^2}\left\langle\left. V^{(1)}(wz_1,\bar{w}\bar{z}_1)V^{(2)}(wz_2,\bar{w}\bar{z}_2)\lp b_0+\widetilde{b}_0\rp\right|B\right\rangle.
\ee

Let's consider the situation in which $|z_1|>|z_2|$ and $V^{(1)}$ is BRST-exact,
\be
V^{(1)}(z,\bar{z})=\left\{Q,\Lambda(z,\bar{z})\right\},
\ee
where $\Lambda(z,\bar{z})$ is a local operator of weight $(0,0)$ and total ghost number one.  We can then compute
\begin{multline}
\int_{|w|>1/|z_2|}\frac{d^2w}{|w|^2}\left\langle\left\{Q,\Lambda(wz_1,\bar{w}\bar{z}_1)\right\}V^{(2)}(wz_2,\bar{w}\bar{z}_2)\lp b_0+\widetilde{b}_0\rp|B\right\rangle\\
=\int_{|w|>1/|z_2|}\frac{d^2w}{|w|^2}\left\langle\Lambda(wz_1,\bar{w}\bar{z}_1)V^{(2)}(wz_2,\bar{w}\bar{z}_2)\lp L_0+\widetilde{L}_0\rp|B\right\rangle\\
=-\int_{|w|>1/|z_2|}\frac{d^2w}{|w|^2}\lp w\frac{\p}{\p w}+\bar{w}\frac{\p}{\p\bar{w}}\rp\left\langle\Lambda(wz_1,\bar{w}\bar{z}_1)V^{(2)}(wz_2,\bar{w}\bar{z}_2)|B\right\rangle,
\end{multline}
where we used the fact that, for an operator of weight $(0,0)$,
\be
\ls L_0+\widetilde{L}_0,\mathcal{O}(wz,\bar{w}\bar{z})\rs=\lp wz\p+\bar{w}\bar{z}\bar{\p}\rp\mathcal{O}(wz,\bar{w}\bar{z})=\lp w\frac{\p}{\p w}+\bar{w}\frac{\p}{\p\bar{w}}\rp\mathcal{O}(wz,\bar{w}\bar{z}).
\ee
We also used the fact that
\be
\left\{Q,b_0+\widetilde{b}_0\right\}=L_0+\widetilde{L}_0,
\ee
and that the boundary state is annihilated by the total BRST charge $Q$.  If we now switch to polar coordinates, $w=re^{i\theta}/z_2$, then we have
\begin{multline}
\label{eq:TwoPointBRSTExact}
2i\int_0^{2\pi}d\theta\int_1^\infty dr\frac{\p}{\p r}\left\langle\Lambda(re^{i\theta}z_1/z_2,re^{-i\theta}\overline{z_1/z_2})V^{(2)}(re^{i\theta},re^{-i\theta})|B\right\rangle\\
=2i\int_0^{2\pi}d\theta\left\langle\Lambda(re^{i\theta}z_1/z_2,re^{-i\theta}\overline{z_1/z_2})V^{(2)}(re^{i\theta},re^{-i\theta})|B\right\rangle\Big|^{r=\infty}_{r=1}.
\end{multline}

If $|z_1|>|z_2|$ but $V^{(2)}$ is the BRST-exact operator, we get a similar expression,
\be
2i\int_0^{2\pi}d\theta\left\langle V^{(1)}(re^{i\theta}z_1/z_2,re^{-i\theta}\overline{z_1/z_2})\Lambda(re^{i\theta},re^{-i\theta})|B\right\rangle\Big|^{r=\infty}_{r=1}.
\ee

In order to argue that BRST-exact states decouple from the disc amplitude then, we need to argue that the boundary contributions above vanish, and this will require an additional assumption on the operators appearing in the amplitude.  In particular, we will assume that $V^{(1)}$ and $V^{(2)}$ carry momentum $p_1$ and $p_2$ respectively, i.e.\ $V^{(1)}$ is constructed from $e^{ip_1X(z_1,\bar{z}_1)}$ multiplied by fields which do not depend on the constant mode of $X^\m(z,\bar{z})$, and similarly $V^{(2)}$ is constructed with $e^{ip_2X}$.  In this familiar situation, the amplitude produces a delta function enforcing momentum conservation along the brane,
\be
\d^{p+1}(p_1+Dp_1+p_2+Dp_2),
\ee
and thus on-shell the only invariants which can be constructed from these momenta are
\be
s=p_1Dp_1,\qquad t=p_1p_2.
\ee
The remaining combinations can be expressed in terms of $s$ and $t$ as
\be
p_1Dp_2=-s-t,\qquad p_2Dp_2=s.
\ee

In addition to the delta function, the OPEs of the exponential factors amongst themselves produces a universal factor
\be
\mathcal{K}=\lp|z_1|^2-1\rp^s\lp|z_2|^2-1\rp^s\left|z_1-z_2\right|^{2t}\left|z_1\bar{z}_2-1\right|^{-2s-2t}.
\ee
All the remaining OPEs in the amplitude will combine to multiply $\mathcal{K}$ by a rational function of the insertion positions, i.e.\ the ratio of two polynomials\footnote{Individual OPEs in R sectors can produce factors with half-integral exponents, but if all the vertex operators satisfy the GSO projection, then the final OPE result will involve only integer exponents.} in $z_1$, $z_2$, $\bar{z}_1$, and $\bar{z}_2$.

If we plug these results into the amplitude (\ref{eq:TwoPoint}), and use variables of integration where the range of the radial integral is from one to infinity, 
then near $r=1$ we have approximately (for some integer $n$ determined by the other contractions in the amplitude)
\be
\sim\int_1dr\lp r-1\rp^{s+n},
\ee
which only converges for $\operatorname{Re}(s)>-n-1$.  At the other end of the range we have (for some integer $m$)
\be
\sim\int^\infty dr\,r^{-2t+m},
\ee
which only converges for $\operatorname{Re}(t)>(m+1)/2$.  Outside of this range of $s$ and $t$ the integral diverges and we cannot make sense of our usual expression.  Fortunately, we expect the physical amplitude to be an analytic function of the momenta $p_1$ and $p_2$.  Thus we can first complexify the momenta and then continue to the region where the integral converges; once we find the answer we're looking for, we can extrapolate back to the region where the integral failed to converge.  Note that with complex momenta we can perform the continuation while remaining on shell.  For instance, with a Lorentzian brane we have $s=p_1^2-2p_1^ip_1^i$, where $i$ indexes the transverse directions.  We can stay on-shell, $p_1^2=0$, and send $s$ to be something with sufficiently positive real part, as long as we let $p_1^i$ have sufficiently large imaginary parts.

The same results will hold for the boundary terms of (\ref{eq:TwoPointBRSTExact}); they will be given by either
\be
\sim\lp r-1\rp^{s+n'}|_{r=1},\qquad\mathrm{or}\qquad\sim r^{-2t+m'}|_{r=\infty}.
\ee
Since we expect the result to be analytic in $s$ and $t$, and since the result is clearly zero if $s$ and $t$ have sufficiently large real parts, it follows that the result must be zero identically.

Let us comment briefly on a couple of situations where this line of reasoning is not available.  If either state is prevented from carrying generic momentum (it could happen that the spectrum of physical states gets a enhanced at zero-momentum for example) then we can't use the argument as presented.  There are also situations where the vertex operators can carry generic momenta, and the amplitude is still expected to be an analytic function of those momenta, but where some momentum invariant vanishes for kinematic reasons.  For example,  in the case of a D9-brane, momentum conservation and the on-shell condition forces $s=t=0$.  And in the case of a D-instanton we can have arbitrary $t$ (after continuation), but $s=-p_1^2=0$.  In these situations it may be that the integral diverges for all on-shell momenta, regardless of how we try to analytically continue.  We shall not discuss these cases further in the present work.

To finish off this section, we shall show that, as a corollary to the decoupling of generic BRST-exact states, the two-point function is independent of the distribution of picture charge.  In the RNS formalism, picture changing is implemented by $X_0$ (or $\widetilde{X}_0$), which is the zero mode of the local operator
\bea
X(z) &=& \sum_{n\in\IZ}X_nz^{-n}=\left\{Q,2\xi(z)\right\}\\
&=& :\lp 2c\p\xi+e^\phi\psi^\m\p X_\m-\hlf\p b\eta e^{2\phi}-b\p\eta e^{2\phi}-b\eta\p\phi e^{2\phi}\rp(z):.\non
\eea
Note that $X(z)$ is not BRST-exact because the field $\xi(z)$ is not included in our algebra of free fields, but $\p X(z)=\left\{Q,2\p\xi(z)\right\}$ {\it{is}} BRST-exact.  Given a BRST-closed physical state $V_P(z,\bar{z})$ of left-moving picture $P$, we define
\be
V_{P+1}(z,\bar{z})=X_0V_P(z,\bar{z})=\oint_{|w|=|z|+\e}\frac{dw}{2\pi iw}X(w)V_P(z,\bar{z}),
\ee
for some small $\e>0$.  It can be verified that $V_{P+1}$ has picture $P+1$, has the correct weight and ghost number to be a physical state, is BRST-closed, and is BRST-exact if and only if\footnote{This statement should only hold at generic momenta~\cite{Berkovits:1997mc}, which is of course the case we are interested in here.  We would like to thank Nathan Berkovits for useful correspondence on this topic.} $V_P$ is BRST-exact.  Thus to redistribute the left-moving picture charge in an amplitude, whether on the sphere or on the disc, we have simply to repeatedly commute copies of $X_0$ through operators to move it to the position we want.

Each time we commute $X_0$ through one of our operators, we pick up a contribution
\begin{multline}
\ls X_0,V_P(z,\bar{z})\rs=\oint_{|w-z|<\e}\frac{dw}{2\pi iw}X(w)V(z)\\
=\oint_{|w-z|<\e}\frac{dw}{2\pi iw}\lp X(w_0)+\int_{w_0}^wdu\p X(u)\rp V_P(z,\bar{z})\\
=\left\{Q,2\oint_{|w-z|<\e}\frac{dw}{2\pi iw}\int_{w_0}^wdu\p\xi(u)V_P(z,\bar{z})\right\}\\
=\left\{Q,2\oint_{|w-z|<\e}\frac{dw}{2\pi iw}\xi(w)V_P(z,\bar{z})\right\}.
\end{multline}
Similar expressions can be derived for the lowering operator, and the right-moving operators.

Since this shows that the additional contribution from redistributing the picture charge is BRST-exact, and since we have already shown that BRST-exact states decouple for generic momenta, we see that we can freely move the picture charge around as long as we keep the total left- and right-picture charges fixed.

Finally, we would like to argue that on the disc we can also move picture charge from the left to the right.  But in fact this is also straight-forward, and requires only the facts that
\be
\ls X_0,b_0+\widetilde{b}_0\rs=\ls\widetilde{X}_0,b_0+\widetilde{b}_0\rs=0,\qquad{\mathrm{and}}\qquad X_0\left|B\right\rangle=\widetilde{X}_0\left|B\right\rangle.
\ee
With these we see that we can always move the operator $X_0$ to the right, generating BRST-exact terms along the way, convert it into $\widetilde{X}_0$ when it hits the boundary state, and then move it to the left to the desired position, again possibly generating BRST-exact states as it commutes back through the operators.  Since under our assumptions all those BRST-exact states decouple, we can freely redistribute picture charge however we like, as long as we keep the total charge constant at $-2$.

\section{Two-point Functions on the Disc}
\label{sec:TwoPointAmplitudes}

In this section we will compute the disc amplitudes corresponding to two closed strings interacting with the D-brane using the formalism that we have developed in the previous sections, paying close attention to the steps needed to compare different pictures and to reconstruct the effective action on the brane.  Our results agree with earlier computations in the literature~\cite{Gubser:1996wt,Garousi:1996ad,Hashimoto:1996kf,Hashimoto:1996bf}.

\subsection{Two NS-NS fields}

To begin, let us start with two NS-NS fields with polarization tensors $\e_{1\,\m\n}$ and $\e_{2\,\m\n}$ and momenta $p_1$ and $p_2$,
which we take to be physical on-shell states.
All the possible combinations of momenta can be expressed in terms of two invariants,
\begin{equation}
s=p_1Dp_1\qquad {\rm and } \qquad t=p_1p_2.
\end{equation}

Putting the first operator in the $(-1,-1)$-picture and the second one in the $(0,0)$-picture, the amplitude is
\begin{equation}
\label{eq:TwoPointFunction}
\langle V_{-1,-1} V_{0,0} \rangle = {\cal N}\frac{\G(1+s)\G(1+t)}{\G(1+s+t)}\left(a_1 + \frac{a_2}{s} + \frac{a_3}{t} + a_4\frac{s}{t} + a_5 \frac{t}{s}  \right).
\end{equation}
Here we have use the result that
\be
\int_{|w|^2>1}d^2w\left|w\right|^a\lp 1-\frac{1}{|w|^2}\rp^b=\pi\frac{\G(-1-\frac{a}{2})\G(1+b)}{\G(-\frac{a}{2}+b)},
\ee
which can be easily obtained using polar coordinates.
The overall normalization is
\begin{equation}
{\cal N} = i\lp 2\pi\rp^{p+2}\d^{p+1}\left(  p_1^a+p_2^a \right),
\end{equation}
and the explicit expressions for the coefficients $a_i$ are
\begin{equation}\label{eq:coe}
\begin{split}
a_1 & = -\Tr (\e_1 D)\Tr (\e_2 D)+\Tr (\e_1 D \e_2 D) -\Tr (\e_1 \e_2^T)\\
a_2 & =  \Tr(\e_1 D)p_1 \left(D\e_2D -  \e_2  \right)p_1 +Dp_1\Big[\e_1 D \e_2 -\half \e_1^T \e_2 -\half \e_1 \e_2^T \Big] Dp_2 +(1 \leftrightarrow 2)   \\
a_3 & = \Big[-(p_1 \e_2 p_1) \Tr(\e_1 D) +p_2 \e_1 D \e_2 p_1 +(1 \leftrightarrow 2)\Big] +Dp_1 \Big[\e_2 \e_1 ^T +\e_2^T \e_1 -(1 \leftrightarrow 2 )\Big]Dp_2  \\
a_4 & = -\Tr (\e_1 \e_2^T)\\
a_5 & =-\Tr (\e_1 D) \Tr (\e_2 D).
\end{split}
\end{equation}
This is the result for the two point function of two NS-NS fields with arbitrary polarizations. This result is obviously symmetric under the interchange of the two NS-NS operators.
To facilitate the comparison with supergravity
we will quote next how \C{eq:TwoPointFunction} simplifies for different polarizations which can be anti-symmetric $\e_{\m\n}^{(B)}= B_{\m\n}$, symmetric traceless $\e_{\m\n}^{(h)} = h_{\m\n}$ or pure trace $\e^{(\P)} ={\P \over \sqrt{8}} (\eta_{\m\n} - l_\m p_\n -l_\n p_\m) $, with $l_\m p^\mu=1$.

The coefficients \C{eq:coe}
vanish if one of the polarizations is symmetric and the other anti-symmetric.
As a result the only non-vanishing amplitudes are $\langle \P \P \rangle$, $\langle \P h \rangle$, $\langle hh  \rangle$ and $\langle BB \rangle$. Explicitly
\begin{equation}
\label{eq:NSFieldTwoPoint}
\begin{split}
\langle \Phi \Phi \rangle  = & {\cal N} \Phi_1 \P_2 \left[ {s \over t} +\half (p-3)^2 \left( 1 + {t \over s} \right) \right] \\
\langle \Phi h \rangle  = & {\cal N} {(p-3) \over \sqrt{2}} \P \left[ 2 {h^a}_a+{4 \over s} h_{a i} p_{ 1}^a p_{ 1}^i +{1\over t}
\left(h_{ab} p_{1}^a p_{1}^b +2 h_{ai} p_{1}^a p_{1 }^i +h_{ij} p_{1}^i p_{1 }^j  \right) + {2 t \over s}  {h^a}_a  \right] , \\
\langle h h \rangle  = & {\cal N} \Big\{4 h_{1a}^ah_{2b}^b + 4 h_1^{ai} h_{2ai}+{8\over s} \left(h_{1a}^a h_{2bi}p_1^b p_1^i -h_{1ai} h_{2b}^b p_1^a p_2^i -2 h_{1a}^i h_{2bi} p_1^a p_1^b\right) \\
& +{s \over t} \left( h_1^{ab}h_{2ab} + 2 h_1^{ai} h_{2ai} + h_1^{ij} h_{2 ij} \right)+ {1\over t} \Big[2h_{1a}^a \left(h_{2bc} p_1^b p_1^c + 2 h_{2bi} p_1^b p_1^i + h_{2 ij} p_1^i p_1^j  \right)   \\
& + 2 \left(h_{1ab} p_1^a p_1^b - 2 h_{1ai} p_1^a p_2 ^i + h_{1ij} p_2^i p_2^j  \right)h_{2c}^c -4h_{1b}^a h_{2ac} p_1^b p_1^c -4 h_{1i}^a h_{2aj} p_1^jp_2^i \\&
-8 h_{1a}^i h_{2bi} p_1^a p_1^b -4h_{1a}^i h_{2ij} p_1^a p_1^j +4 h_{1j}^ih_{2ai}p_1^a p_2^j\Big]+ 4 { t \over s} h_{1a}^a h_{2b } ^b\Big\}, \\
\langle B B \rangle  = & {\cal N} \Big[ 2 B_1^{ab} B_{2ab}+ 2 B_1^{ij} B_{2 ij}-{16\over s}  B_{1b}^a B_{2ac} p_{1}^b p_{1}^c+{s \over t}
\left(B_1^{ab} B_{2ab} + 2 B_1^{ai} B_{2ai} +B_1^{ij} B_{2 ij}  \right)     \\
& -{4\over t} \left(2B_{1b}^{a} B_{2ac} p_{1}^b p_{1}^c + B_{1b}^a B_{2ai} p_{1}^b p_{1}^i - B_{1i}^{a} B_{2ab}p_{1}^b p_{2}^i \right.\\
& \left.+ B_{1ai} {B_{2b} }^i p_{1}^a p_{1}^b + B_{1j}^{i} B_{2ik} p_{1}^k p_{2}^j \right)\Big] \\
\end{split}
\end{equation}
In section \ref{subsec:NSNSSUGRA}, we will describe in detail
the supergravity interpretation of the $\langle \P \P \rangle$ correlator, and the comparison for the others appears in appendix \ref{subsec:NSNSAppendix}. We will pay particular attention to the
origin of the poles in the $s$ and $t$ parameters.

Since the fields above must be physical on-shell states, we need $p_1^2=p_2^2=0$ and $\e_{1\,\m\n}p_1^\n=\e_{1\,\n\m}p_1^\n=\e_{2\,\m\n}p_2^\n=\e_{2\,\n\m}p_2^\n=0$, and we also have momentum conservation, $(p_1+p_2)^a=0$.  To eliminate the ambiguities due to these conditions we have made systematically the following replacements,
\bea
p_2^a &=& -p_1^a,\non\\
\e_{1\,\m i}p_1^i &=& -\e_{1\,\m a}p_1^a,\non\\
\e_{1\,i\m}p_1^i &=& -\e_{1\,a\m}p_1^a,\\
\e_{2\,\m i}p_2^i &=& \e_{2\,\m a}p_1^a,\non\\
\e_{2\,i\m}p_2^i &=& \e_{2\,a\m}p_1^a.\non
\eea

\subsection{One R-R and one NS-NS field}
\label{subsec:RRNSDisc}

We will consider the vertex operators with pictures (-1/2,-1/2) and (-1,0), first. Then we will consider the pictures
(-1/2,-1/2) and (0,-1) and verify explicitly that the results are independent of the choice of picture. In section \ref{sec:ComparisonToSUGRA} we will show how to obtain the result from the string amplitude from interactions in space-time and on the D-brane.
The amplitude is
\begin{multline}
\label{eq:(-1,0)Raw}
\langle V_{-\half , -\half } V_{-1,0} \rangle =
\frac{i\pi}{2\sqrt{2}}\frac{\G(t+1)\G(s+1)}{\G(t+s+1)}
\Big\{-\frac{1}{t}\lp\e D\rp_{\m\n}\lp Dp_2\rp_\rho T^{\m\n\rho}+\\
\left[-\frac{2}{s}\lp\e Dp_2\rp_\m+\frac{2}{t}\lp\e p_1\rp_\m -
{ 2 r \over s t} \left[\tr\lp\e D\rp\lp Dp_2\rp_\m-\lp p_2D\e D\rp_\m\right]\right] T^\m\Big\}
\end{multline}
The traces $T^{\m\cdots}$ are given in section \ref{subsec:Traces}, and we have defined the useful combination
\be
r=t+\frac{s}{2}.
\ee

Now let us try to reduce this to the on-shell results.
We will list the contributions according to the degree of the RR field involved:

\begin{itemize}

\item Consider first $F^{(p-2)}$.  The only possible term would come from the $T^{\m\n\rho}$ term above and would be proportional to
\be
\e^{a_1\cdots a_{p+1}} \e_{a_1 a_2}p_{2\, a_3} F^{(p-2)} _{a_4\cdots a_{p+1}},
\ee
but this is zero since $p_{2\, a_3}=-p_{1\, a_3}$, and antisymmetrizing with $F$ then gives $dF\w\e=0$, since $F$ is closed.

\item Next we have $F^{(p)}$.  In order to make full use of the on-shell and physical state conditions,
we will split all indices into along the brane and transverse to the brane, and we need to agree to
always make certain substitutions:
\bea
\label{eq:SubRules}
p_{2\,a} &\rightarrow& -p_{1\,a},\non\\
\e_{\m i}p_2^i &\rightarrow& \e_{\m a}p_1^a,\non\\
\e_{i\m} p_2^i &\rightarrow& \e_{a\m}p_1^a,\\
C_{\m_1\cdots\m_ni}p_1^i &\rightarrow& -C_{\m_1\cdots\m_na}p_1^a.\non
\eea
Also, whenever we have indices of $C$ which are along the brane and which are not contracted by the volume form of the brane (like the index $a$ on the right hand side of the bottom line of (\ref{eq:SubRules})), then we will rewrite $C$ using substitutions like
\be
\label{eq:CSubRule}
\frac{1}{(p-2)!}\e^{a_1\cdots a_{p+1}}C_{a_1\cdots a_{p-2}b}=\frac{3}{(p-1)!}\e^{c_1\cdots c_{p+1}}C_{c_1\cdots c_{p-1}}\d^{[a_{p-1}}_b\d^{a_p}_{c_p}\d^{a_{p+1}]}_{c_{p+1}}.
\ee
The possible contractions of momenta in this scheme are also quite constrained,
\be
p_1^ap_{1\,a}=-p_1^ip_{1\,i}=-p_1^ap_{2\,a}=p_2^ap_{2\,a}=-p_2^ip_{2\,i}=\frac{s}{2},\qquad p_1^ip_{2\,i}=r.
\ee
Employing all these substitutions
\begin{equation}
\begin{split}
\label{eq:(-1,0)C(p-1)}
\Big[ &  \frac{1 }{(p-1)!} \Big( {r\over t}   \e^{a_1a_2}-{ 8 r \over s t }
{\e^{a_1}}_ b p_1^b p_1^{a_2}  - {2\over t}  {\e^{a_1}}_i p_1^i p_1^{a_2 }\Big)  C^{a_3\cdots a_{p+1}} + \\
 &   {(-1)^{p+1}\over (p-2)!}{1\over t}  \e^{a_1 a_2 }p_1^{a_3}p_2^i  {C^{a_4\cdots a_{p+1}}}_i  \Big] \e_{a_1\cdots a_{p+1}}.\\
\end{split}
\end{equation}
In this case $\e$ is anti-symmetric and
the result vanishes if $\e$ is symmetric; only the $B$-field interacts with $C^{(p-1)}$, as expected.

\item Next we turn to $F^{(p+2)}$.  Following the same procedure we find a result proportional to
\begin{equation}
\label{eq:RRSymmetricNSNS}
\begin{split}
&\Big\{  {(-1)^{p+1} \over (p+1)!} \Big[{  r^2 \over s t}\left( \e^b_{\hphantom{b}b}-\e^j_{\hphantom{j}j}\right)+
{ 4 r \over s t}\e_{bi} p_1^b p_1^i  +{1 \over t} \e_{ij} p_1^i p_1^j   \Big] C^{a_1 \cdots a_{p+1}}+\\ &
 {1\over p!} \Big[ {  r \over s t} p_1^{a_1} p_2^i \left( \e^b_{\hphantom{b}b}-\e^j_{\hphantom{j}j}\right) +
{ 4 r \over s t}{\e_b}^i p_1^{a_1} p_1^b  - {  r \over t} {\e^{a_1 i}} +{1\over t} {\e^{a_1}}_ j
p_1^j p_2^i + {1\over t} {\e^i}_j p_1^{a_1}  p_1^j \Big]{C^{a_2 \cdots a_{p+1}}}_i+\\ &
 {(-1)^p\over (p-1)!}
{1 \over t}  \e^{a_1 i} p_1^{a_2} p_2 ^j {C^{a_3 \cdots a_{p+1}}}_{ij} \Big\} \e_{a_1\cdots a_{p+1}} .
\end{split}
\end{equation}
This time only symmetric polarizations (graviton and dilaton) can contribute.

\item The remaining couplings involve $F^{(p+4)}$ and are necessarily of the form
\be
\label{eq:F(p+4)Coupling}
\frac{1}{(p+1)!}\frac{1}{t}\e^{a_1\cdots a_{p+1}}F_{a_1\cdots a_{p+1}ijk}\e^{ij}p_2^k.
\ee

\end{itemize}

We would like to compare this to a computation done in picture (-1/2,-1/2) and (0,-1). In this picture the amplitude is
\begin{multline}
\label{eq:(0,-1)Raw}
\langle V_{-\half , -\half } V_{0,-1} \rangle =
\frac{i\pi}{2\sqrt{2}}\frac{\G(t+1)\G(s+1)}{\G(t+s+1)}
\Big\{\frac{1}{t}\lp\e D\rp_{\m\n}p_{2\,\rho}T^{\m\n\rho}+\\
\left[\frac{2}{s}\lp p_2 D \e D\rp_\m-\frac{2}{t}\lp p_1\e D\rp_\m +
{ 2 r \over s t }\left[\tr\lp\e D\rp p_{2\,\m}-\lp\e D p_2  \rp_\m\right]\right] T^\m\Big\}
\end{multline}
There are a couple of ways we can compare this amplitude to (\ref{eq:(-1,0)Raw}).  We could separate all of the indices into tangent or normal to the brane world-volume, and then make use of the rules (\ref{eq:SubRules}) to see that we do indeed get identical results for on-shell amplitudes.  There is also a more direct comparison which is worth sketching out, however.  There is a certain $\zet_2$ symmetry enjoyed by the string world-sheet theory with boundary conditions given by our D$p$-brane.  This is the symmetry which acts by world-sheet parity $\Omega$, reflection $\s_{9-p}$ in the space-time directions normal to the brane, and for certain values of $p$ carries an additional sign for left-moving space-time fermion number, $(-1)^{F_L}$.  Explicitly the generator is given by
\be
g=\left\{\begin{matrix}\Omega\s_{9-p}\lp -1\rp^{F_L}, & \mathrm{for\ }p=-1,2,3,6,7, \\ \Omega\s_{9-p}, & \mathrm{for\ }p=0,1,4,5,8,9.\end{matrix}\right.
\ee
If we were to quotient the theory by this symmetry, we would be effectively creating an O$p$-plane on top of the D$p$-brane.  Here we don't wish to perform the quotient, but wish to use the fact that $g$ acts as
\bea
\label{eq:ParityAction}
\lp p_n\rp_\m &\mapsto& \lp Dp_n\rp_\m,\non\\
\e_{\m\n} & \mapsto& D_\m^{\hphantom{\m}\rho}D_\n^{\hphantom{\n}\s}\e_{\s\rho}=\lp D\e^TD\rp_{\m\n},\\
F^{(p+2k)}_{\m_1\cdots\m_{p+2k}} &\mapsto& \lp -1\rp^{k+1}D_{\m_1}^{\hphantom{\m_1}\n_1}\cdots D_{\m_{p+2k}}^{\hphantom{\m_{p+2k}}\n_{p+2k}}F^{(p+2k)}_{\n_1\cdots\n_{p+2k}}.\non
\eea
One can then verify using (\ref{eq:ExplicitTraceFormula}) that $g$ sends
\be
\label{eq:ParityTrace}
T^{\m_1\cdots\m_n}\mapsto\lp -1\rp^{\frac{n+1}{2}}T^{\m_1\cdots\m_n}.
\ee
It is straightforward to then check that under (\ref{eq:ParityAction}) and (\ref{eq:ParityTrace}) the result (\ref{eq:(0,-1)Raw}) is mapped into the result (\ref{eq:(-1,0)Raw}).

It is of course expected that the two pictures agree, in light of the arguments in section \ref{sec:BRSTExact} but it is interesting to see the mechanism.  Let us also consider the same computation in the $(-\frac{3}{2},-\hlf)$-$(0,0)$ picture, where we take the $y=0$ gauge (\ref{eq:y=0Gauge}) for the R-R vertex operator.  Recall that this version of the vertex operator only lived in the absolute cohomology since it was annihilated by $\widetilde{b}_0$ but not by $b_0$.  In order to get a nonzero contribution we must have total $\phi$ and $\wtphi$ charge of $-2$, which means we must take from the NS-NS $(0,0)$ picture operator either the term with $e^\phi$ or the term with $e^\wtphi$.  In each case we can evaluate the ghost and superghost correlators, reducing the amplitude to correlators in the matter sectors alone.  In the latter case, the entire contribution vanishes in the limit $z_1\rightarrow\infty$, while in the former case we are left with
\begin{multline}
\left\langle 0\left|2\lp\mathcal{C}\slashed{F}\rp_{AB}c\p ce^{-\frac{5}{2}\phi}\p\xi S^A\wtc e^{-\hlf\wtphi}\wtS^Be^{ip_1X}(z_1,\barz_1)\right.\right.\\
\left.\left.\times -\hlf\e_{\m\n}e^\phi\eta\psi^\m\wtc\lp\bar{\p}X^\n-ip_\rho\wtpsi^\rho\wtpsi^\n\rp e^{ip_2X}(z_2,\barz_2)\right|B;\eta\right\rangle_R\\
=\lp\mathcal{C}\slashed{F}\rp_{AB}\e_{\m\n}z_2^\hlf\barz_2\left\langle A,B\left|e^{ip_1X}\times\psi^\m\lp\bar{\p}X^\n-ip_\rho\wtpsi^\rho\wtpsi^\n\rp e^{ip_2X}(z_2,\barz_2)\right|B_{X,\psi};\eta\right\rangle_R.
\end{multline}

But this result precisely matches what we get in the $(-\hlf,-\hlf)$-$(-1,0)$ picture computation,
\begin{multline}
\left\langle 0\left|\lp\mathcal{C}\slashed{F}\rp_{AB}ce^{-\hlf\phi}S^A\wtc e^{-\hlf\wtphi}\wtS^Be^{ip_1X}(z_1\barz_1)\right.\right.\\
\left.\left.\times ce^{-\phi}\psi^\m\wtc\lp\bar{\p}X^\n-ip_\rho\wtpsi^\rho\wtpsi^\n\rp e^{ip_2X}(z_2,\barz_2)\right|B;\eta\right\rangle_R\\
=\lp\mathcal{C}\slashed{F}\rp_{AB}\e_{\m\n}z_2^\hlf\barz_2\left\langle A,B\left|e^{ip_1X}\times\psi^\m\lp\bar{\p}X^\n-ip_\rho\wtpsi^\rho\wtpsi^\n\rp e^{ip_2X}(z_2,\barz_2)\right|B_{X,\psi};\eta\right\rangle_R.
\end{multline}
So the two pictures will certainly agree.  Note that this argument easily extends to include any number of additional integrated operators in the $(0,0)$ picture, since these are independent of the ghosts and superghosts.


\section{Comparison with Space-Time Lagrangian}
\label{sec:ComparisonToSUGRA}

\subsection{Two NS-NS fields}
\label{subsec:NSNSSUGRA}

We will now show how the correlators of (\ref{eq:NSFieldTwoPoint}) can be obtained by evaluating field theory diagrams (parts of this computation have appeared before, in~\cite{Gubser:1996wt,Garousi:1996ad,Hashimoto:1996kf,Hashimoto:1996bf}).  We will work to leading order in momenta, meaning that we will take only the leading constant term in the expansion
\be
\frac{\G(1+s)\G(1+t)}{\G(1+s+t)}=1-\frac{\pi^2}{6}st+\zeta(3)st\lp s+t\rp+\cdots.
\ee

\begin{figure}
\label{fig:SUGRADiagrams}
\begin{center}
\scalebox{.5}{\includegraphics{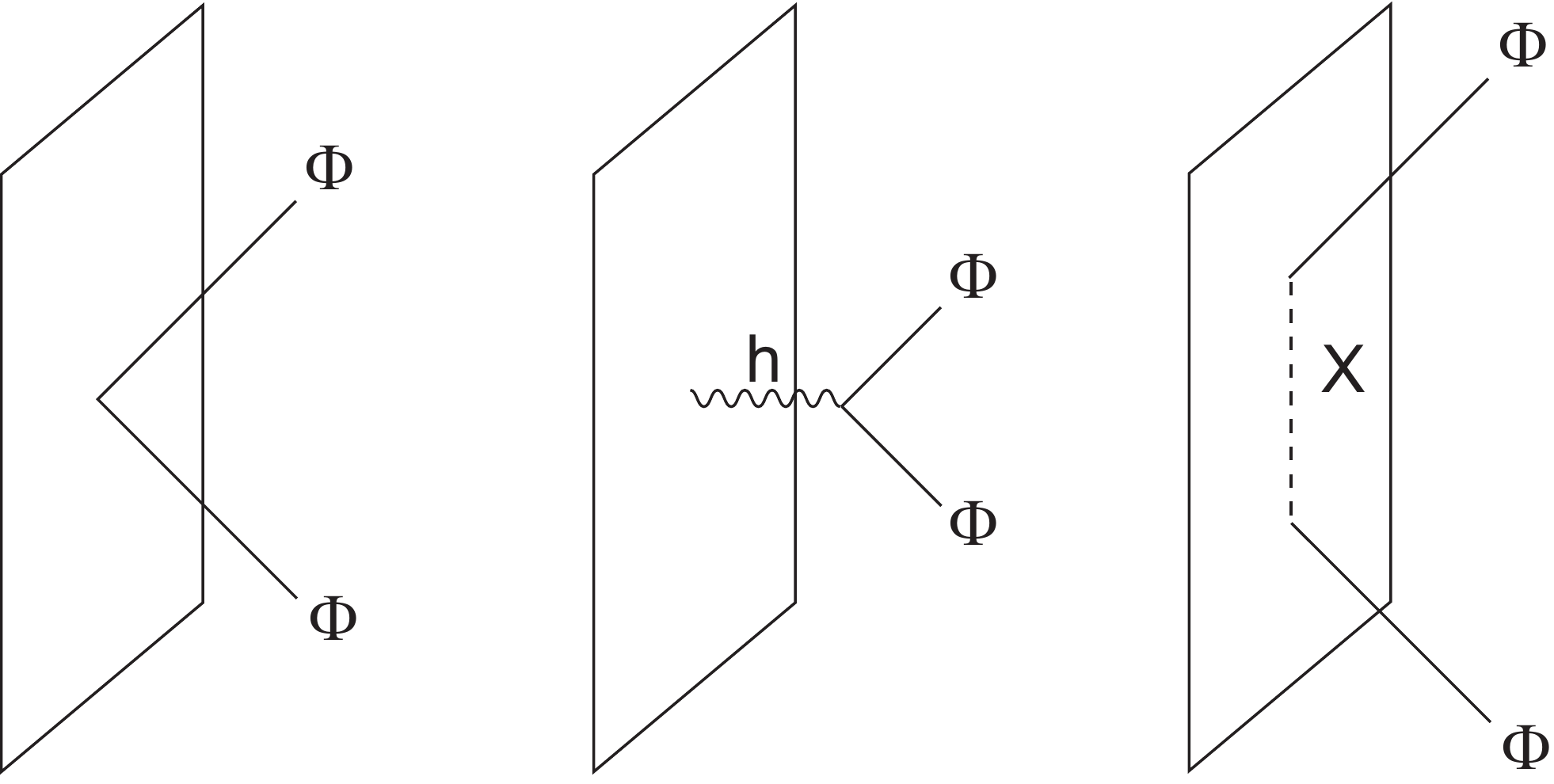}} \caption{Dilaton two point function.}
\end{center}
\end{figure}

Let's begin with $\langle\Phi\Phi\rangle$.  We will show that in supergravity three diagrams contribute to this amplitude, as shown in Figure \ref{fig:SUGRADiagrams}.  These three diagrams correspond to a contact term on the brane, an interaction in the bulk which produces a graviton, which is then absorbed by the brane, or a process where each dilaton hits the brane and they exchange a scalar field on the brane.  These three diagrams are then related to the amplitude by
\begin{multline}
\label{eq:PhiPhiSUGRAAmplitude}
\left\langle\Phi_1\Phi_2\right\rangle\sim\Phi_1\Phi_2\left\{\lp\frac{\d}{\d\Phi_1}\frac{\d}{\d\Phi_2}S_{p+1}\rp+i\lp\frac{\d}{\d\Phi_1}\frac{\d}{\d\Phi_2}\frac{\d}{\d h_{\m\n}}S_{10}\rp G^{(h)}_{\m\n,\rho\s}\lp\frac{\d}{\d h_{\rho\s}}S_{p+1}\rp\right.\\
\left.+i\lp\frac{\d}{\d\Phi_1}\frac{\d}{\d X^i}S_{p+1}\rp G^{(X)\,ij}\lp\frac{\d}{\d\Phi_2}\frac{\d}{\d X^j}S_{p+1}\rp\right\}.
\end{multline}
Here the objects $G^{(h)}$ and $G^{(X)}$ are propagators for the graviton and the scalar field on the brane respectively, and the three terms above represent the three diagrams of the figure.

We will be working with fields whose kinetic terms are canonically normalized, so it is easy to list the propagators of all the fields we need in the bulk
\bea
\label{eq:NSNSPropagators}
G^{(\Phi)} &=& \frac{-i}{p^2},\non\\
G^{(h)}_{\m\n,\rho\s} &=& \frac{-i}{2p^2}\lp\eta_{\m\rho}\eta_{\n\s}+\eta_{\m\s}\eta_{\n\rho}-\frac{1}{4}\eta_{\m\n}\eta_{\rho\s}\rp,\non\\
G^{(B)}_{\m\n,\rho\s} &=& \frac{-i}{2p^2}\lp\eta_{\m\rho}\eta_{\n\s}-\eta_{\m\s}\eta_{\n\rho}\rp,\\
\eea
and on the brane
\bea
G^{(A)}_{ab} &=& \frac{-i}{q^2}\eta_{ab},\non\\
G^{(X)\,ij} &=& \frac{-i}{q^2}\d^{ij},
\eea
where the momentum $q_a$ lies only along the brane.  Here we have chosen specific gauges for the propagators $G^{(h)}$, $G^{(B)}$, and $G^{(A)}$, but the physical amplitude should not depend on these choices.  We will mention some useful relations for the graviton propagator,
\be
G^{(h)}_{ab,cd}\eta^{cd}=i\frac{p-7}{16t}\eta_{ab},\qquad G^{(h)}_{ai,bc}\eta^{bc}=0,\qquad G^{(h)}_{ij,ab}\eta^{ab}=i\frac{p+1}{16t}\d_{ij}.
\ee

Now we need to convert the NS-NS sector of the bulk action and the DBI action on the brane into forms where all fluctuating fields are canonically normalized.  In string frame, the bulk action is given by
\be
S_{10}=\frac{1}{2\k^2}\int d^{10}x\sqrt{-G_s}\,e^{-2\Phi_s}\lp R_s+4\p^\m\Phi_s\p_\m\Phi_s-\frac{1}{12}H_s^{\m\n\rho}H_{s\,\m\n\rho}\rp,
\ee
where we use a subscript s for string frame.  To use the propagators above, we need to change variables as follows,
\be
G_{s\,\m\n}=e^{\Phi_s/2}G_{\m\n},\qquad\Phi_s=\sqrt{2}\k\Phi,\qquad B_s=2\k B,
\ee
which converts the action to
\be
S_{10}=\int d^{10}x\sqrt{-G}\lp\frac{1}{2\k^2}R-\hlf\p^\m\Phi\p_\m\Phi-\frac{1}{6}e^{-\sqrt{2}\k\Phi}H^{\m\n\rho}H_{\m\n\rho}\rp.
\ee
If additionally we write
\be
G_{\m\n}=\eta_{\m\n}+2\k h_{\m\n},
\ee
then $\Phi$, $h_{\m\n}$, $B_{\m\n}$ are all canonically normalized and have the propagators (\ref{eq:NSNSPropagators}).

The DBI action in string frame is
\bea
S^{(DBI)}_{p+1} &=& -\m_p\int d^{p+1}x e^{-\Phi_s}\ls-\det\lp g_s+B_s+2\pi\alpha'F_s\rp\rs^{1/2}\\
&=& -\m_p\int d^{p+1}x e^{\frac{p-3}{2\sqrt{2}}\k\Phi}\ls-\det\lp g+2\k e^{-\frac{1}{\sqrt{2}}\k\Phi}B+\frac{1}{\sqrt{\m_p}}e^{-\frac{1}{\sqrt{2}}\k\Phi}F\rp\rs^{1/2}.\non
\eea
Here we have switched to a canonically normalized gauge field.  We will also use $X^i_s=X^i/\sqrt{\m_p}$ for the same reason.  We work in a gauge in which $X^a(x)=x^a$ (where $X^\m$ are the scalars describing the embedding of the brane into space-time, and $x^a$ are the coordinates on the world-volume), so only $X^i(x)$ are propagating fields.  The bulk fields have been pulled back to the brane, so for example we use (up to rescaling the $X^i$)
\be
\Phi=\Phi(X)=\Phi(x)+X^i\p_i\Phi(x)+\hlf X^iX^j\p_i\p_j\Phi(x)+\cdots,
\ee
and
\begin{multline}
g_{ab}=G_{ab}(X)+G_{ai}(X)\p_bX^i(x)+G_{bi}(X)\p_aX^i(x)+G_{ij}(X)\p_aX^i(x)\p_bX^j(x)\\
=\eta_{ab}+\p_aX^i\p_bX_i+2\k\lp\vphantom{\hlf}h_{ab}+X^i\p_ih_{ab}+h_{ai}\p_bX^i+h_{bi}\p_aX^i\right.\\
\left.+\hlf X^iX^j\p_i\p_jh_{ab}+X^j\p_jh_{ai}\p_bX^i+X^j\p_jh_{bi}\p_aX^i+h_{ij}\p_aX^i\p_bX^j+\cdots\rp.
\end{multline}

Expanding $S_{p+1}$ to the order we need, we find
\begin{multline}
S_{p+1}=\int d^{p+1}x\left\{-\hlf\p^aX^i\p_aX_i-\frac{1}{4}F^{ab}F_{ab}-\m_p\k\lp\frac{p-3}{2\sqrt{2}}\Phi+h^a_{\hphantom{a}a}\rp\right.\\
\left.-\m_p\k^2\lp\frac{(p-3)^2}{16}\Phi^2+\frac{p-3}{2\sqrt{2}}\Phi h^a_{\hphantom{a}a}+\hlf h^a_{\hphantom{a}a}h^b_{\hphantom{b}b}-h^{ab}h_{ab}+B^{ab}B_{ab}\rp\right.\\
\left.-\sqrt{\m_p}\k\lp\frac{p-3}{2\sqrt{2}}\p^i\Phi X_i+\p^ih^a_{\hphantom{a}a}X_i+2h^{ai}\p_aX_i+2B^{ab}\p_aA_b\rp+\cdots\right\}.
\end{multline}

From these actions we can derive the variations we need\footnote{We are ignoring the delta functions enforcing momentum conservation which also come from these variations, and which would be identical to the delta functions which emerge from the disc amplitudes; these would be easy to restore, but since we did not carefully keep track of the overall normalization constant of the disc amplitudes, this restoration would not gain us anything.}
\bea
\frac{\d}{\d\Phi_1}\frac{\d}{\d\Phi_2}S_{p+1} &=& -\m_p\k^2\frac{\lp p-3\rp^2}{8},\non\\
\frac{\d}{\d\Phi_1}\frac{\d}{\d\Phi_2}\frac{\d}{\d h_{\m\n}}S_{10} &=& \k t\eta^{\m\n}-\k\lp p_1^\m p_2^\n+p_1^\n p_2^\m\rp,\\
\frac{\d}{\d h_{ab}}S_{p+1} &=& -\m_p\k\eta^{ab},\qquad\frac{\d}{\d h_{ai}}S_{p+1}=\frac{\d}{\d h_{ij}}S_{p+1}=0,\non\\
\frac{\d}{\d\Phi_1}\frac{\d}{\d X^i}S_{p+1} &=& -i\sqrt{\m_p}\k\frac{p-3}{2\sqrt{2}}p_{1\,i}.\non
\eea
We have assumed that $\Phi_1$ and $\Phi_2$ are on-shell states with momenta $p_1$ and $p_2$ respectively, but we of course do not assume that the fields corresponding to internal lines are on-shell.

Now we can plug these results into (\ref{eq:PhiPhiSUGRAAmplitude}) to find
\bea
\left\langle\Phi_1\Phi_2\right\rangle &\sim& \Phi_1\Phi_2\left\{-\m_p\k^2\frac{(p-3)^2}{8}+i\lp\k t\eta^{\m\n}-\k\lp p_1^\m p_2^\n+p_1^\n p_2^\m\rp\rp G^{(h)}_{\m\n,ab}\lp-\m_p\k\eta^{ab}\rp\right.\non\\
&& \qquad\left.+i\lp-i\sqrt{\m_p}\k\frac{p-3}{2\sqrt{2}}p_{1\,i}\rp G^{(X)\,ij}\lp-i\sqrt{\m_p}\k\frac{p-3}{2\sqrt{2}}p_{2\,j}\rp\right\}\non\\
&=& -\m_p\k^2\Phi_1\Phi_2\left\{\frac{(p-3)^2}{8}-\frac{p-7}{16t}\lp\lp p+1\rp t+s\rp-\frac{p+1}{16t}\lp\lp 7-p\rp t-s\rp\right.\non\\
&& \qquad\left.+\frac{(p-3)^2}{8}\frac{t+s/2}{s/2}\rp\\
&=& -\hlf\m_p\k^2\Phi_1\Phi_2\left\{\frac{s}{t}+\frac{(p-3)^2}{2}\lp 1+\frac{t}{s}\rp\right\}.\non
\eea
Up to an overall normalization factor which we have not tried to determine carefully, this is in perfect agreement with (\ref{eq:NSFieldTwoPoint}).

The other three nonvanishing amplitudes can be treated similarly, using (schematically)
\bea
\left\langle\Phi h\right\rangle &\sim& \Phi h\left\{\lp\frac{\d}{\d\Phi}\frac{\d}{\d h}S_{p+1}\rp+i\lp\frac{\d}{\d\Phi}\frac{\d}{\d h}\frac{\d}{\d\Phi}S_{10}\rp G^{(\Phi)}\lp\frac{\d}{\d\Phi}S_{p+1}\rp\right.\non\\
&& \qquad\left.+i\lp\frac{\d}{\d\Phi}\frac{\d}{\d X}S_{p+1}\rp G^{(X)}\lp\frac{\d}{\d h}\frac{\d}{\d X}S_{p+1}\rp\right\},\non\\
\left\langle hh\right\rangle &\sim& h_1h_2\left\{\lp\frac{\d}{\d h_1}\frac{\d}{\d h_2}S_{p+1}\rp+i\lp\frac{\d}{\d h_1}\frac{\d}{\d h_2}\frac{\d}{\d h}S_{10}\rp G^{(h)}\lp\frac{\d}{\d h}S_{p+1}\rp\right.\non\\
&& \qquad\left.+i\lp\frac{\d}{\d h_1}\frac{\d}{\d X}S_{p+1}\rp G^{(X)}\lp\frac{\d}{\d h_2}\frac{\d}{\d X}S_{p+1}\rp\right\},\\
\left\langle BB\right\rangle &\sim& B_1B_2\left\{\lp\frac{\d}{\d B_1}\frac{\d}{\d B_2}S_{p+1}\rp+i\lp\frac{\d}{\d B_1}\frac{\d}{\d B_2}\frac{\d}{\d\Phi}S_{10}\rp G^{(\Phi)}\lp\frac{\d}{\d\Phi}S_{p+1}\rp\right.\non\\
&& \qquad\left.+i\lp\frac{\d}{\d B_1}\frac{\d}{\d B_2}\frac{\d}{\d h}S_{10}\rp G^{(h)}\lp\frac{\d}{\d h}S_{p+1}\rp\right.\non\\
&& \qquad\left.+i\lp\frac{\d}{\d B_1}\frac{\d}{\d A}S_{p+1}\rp G^{(A)}\lp\frac{\d}{\d B_2}\frac{\d}{\d A}S_{p+1}\rp\right\}.\non
\eea
We relegate the details to appendix \ref{subsec:NSNSAppendix}.

\subsection{One R-R and one NS-NS field}

We will need one more propagator, for R-R fields in the bulk,
\be
G^{(C^{(n)})}_{\m_1\cdots\m_n,\n_1\cdots\n_n}=\frac{-i}{n!p^2}\lp\eta_{\m_1\n_1}\cdots\eta_{\m_n\n_n}\pm\mathrm{perms}\rp.
\ee
A useful relation is that
\be
\label{eq:RRPropagatorContraction}
G^{(C^{(p+1)})}_{a_1\cdots a_{p+1},b_1\cdots b_{p+1}}\e^{b_1\cdots b_{p+1}}=\frac{-i}{p^\m p_\m}\e_{a_1\cdots a_{p+1}},\qquad G^{(C^{(p+1)})}_{i\m_1\cdots\m_p,a_1\cdots a_{p+1}}=0.
\ee

The conventional way of writing the bulk action for the R-R fields in either type IIA or IIB is inconvenient for our purposes for several reasons - we use duality to eliminate the higher degree potentials $C_n$, $n>4$, we have to deal with both kinetic terms and Chern-Simons terms for the remaining fields, and in IIB we have to impose the self-duality of $F^{(5)}$ by hand.  There is an alternative formulation which suits our purposes much better and is known as the democratic formulation~\cite{Bergshoeff:2001pv},
\be
S_{10}=-\frac{1}{8\k^2}\int d^{10}x\sqrt{-G_s}\sum_n\left|dC_s^{(n)}+H_s\w C_s^{(n-2)}\right|^2,
\ee
where for an $(n+1)$-form we use the notation
\be
\left|\om_{n+1}\right|^2=\frac{1}{(n+1)!}\om^{\m_1\cdots\m_{n+1}}\om_{\m_1\cdots\m_{n+1}}.
\ee
Notice that there are no Chern-Simons terms when the action is written this way.  However, in this formulation we have to impose the duality constraints by hand.  In principle, when varying the action with respect to one of the R-R potentials, we should first rewrite all occurrences of the dual potential in terms of the one we are interested in, and then take the variation.  In practice, this simply means that we get an extra factor of two from the action above, and we can proceed as if each of our bulk vertices comes from the variation of a term
\be
S_{10}\supset -\frac{1}{4\k^2}\int d^{10}x\sqrt{-G_s}\left|dC_s^{(n)}+H_s\w C_s^{(n-2)}\right|^2
\ee
in the action\footnote{It is easy to check using the conventional action that this procedure works for low-degree potentials, for instance in IIB we have a term
\be
-\frac{1}{4\k^2}\int d^{10}x\sqrt{-G_s}\left|dC_s^{(2)}+C_s^{(0)}H_s\right|^2,
\ee
and for either the $C^{(2)}C^{(2)}h$ or $C^{(2)}C^{(0)}B$ bulk vertices, this is the only contribution.}.

To convert to normalized kinetic terms we need to define
\be
C_{s\,\m_1\cdots\m_n}^{(n)}=\sqrt{2}\k C^{(n)}_{\m_1\cdots\m_n},
\ee
and the action above becomes
\be
S_{10}=-\hlf\int d^{10}x\sqrt{-G}\sum_ne^{\frac{4-n}{\sqrt{2}}\k\Phi}\left|dC^{(n)}+2\k H\w C^{(n-2)}\right|^2.
\ee

Meanwhile, the Wess-Zumino part of the brane action becomes
\begin{multline}
\label{eq:WZActionExpansion}
S_{p+1}=\sqrt{2}\m_p\k\int Ce^{2\k B+\frac{1}{\sqrt{\m_p}}F}\\
=\int d^{p+1}x\e^{a_1\cdots a_{p+1}}\left\{\frac{\sqrt{2}\m_p\k}{(p+1)!}C^{(p+1)}_{a_1\cdots a_{p+1}}+\frac{\sqrt{2}\m_p\k^2}{(p-1)!}C^{(p-1)}_{a_1\cdots a_{p-1}}B_{a_pa_{p+1}}\right.\\
\left.+\frac{\sqrt{2\m_p}\k}{(p+1)!}\p^iC^{(p+1)}_{a_1\cdots a_{p+1}}X_i+\frac{\sqrt{2\m_p}\k}{p!}C^{{\rlap{$\mathsurround=0pt\scriptstyle{(p+1)}$}}\hphantom{a_1\cdots a_p}i}_{a_1\cdots a_p}\p_{a_{p+1}}X_i\right.\\
\left.+\frac{\sqrt{2\m_p}\k}{(p-1)!}C^{(p-1)}_{a_1\cdots a_{p-1}}\p_{a_p}A_{a_{p+1}}+\cdots\right\}.
\end{multline}

With these preliminaries, we can compute the expected contributions to the amplitudes of section \ref{subsec:RRNSDisc}, namely $\langle C^{(p+3)}B\rangle$, $\langle C^{(p-1)}B\rangle$, $\langle C^{(p+1)}\Phi\rangle$, and $\langle C^{(p+1)}h\rangle$.  The computations are straightforward but long, so we again leave the details to an appendix, \ref{subsec:RRAppendix}.  There it can be verified that these field theory computations exactly agree\footnote{This clarifies a confusion regarding the string theory amplitude computation of the $\int C\w B$ coupling mentioned in~\cite{Craps:1998fn,Craps:2000zr}.} with the disc amplitude computations, up to an overall normalization (but, again with the same normalization for all four of the non-vanishing two-point functions).

\acknowledgments

The authors would like to thank Nathan Berkovits, Jacques Distler, Ruben Minasian, Rob Myers, and Niclas Wyllard for useful discussions and comments. This research was supported in part by NSF Grant No. PHY05-55575, NSF
Grant No. PHY09-06222, NSF Gant No. PHY05-51164, Focused Research
Grant DMS-0854930, Texas A{\&}M University, and the Mitchell
Institute for Fundamental Physics and Astronomy.

\appendix

\section{Gamma matrix conventions}
\label{sec:GammaMatrixConventions}

We define flat space gamma matrices $(\G^\m)^A_{\hphantom{A}B}$ which obey
\be
\left\{\G^\m,\G^\n\right\}=2\eta^{\m\n},
\ee
and we write $\G^{\m_1\cdots\m_n}=\G^{[\m_1}\G^{\m_2}\cdots\G^{\m_n]}$ for antisymmetrized products of gamma matrices.  We also define
\be
\G_{11}=\G^{0\cdots 9}=-\frac{1}{10!}\e_{\m_1\cdots\m_{10}}\G^{\m_1\cdots\m_{10}},
\ee
(note that we use $\e^{0\cdots 9}=1$, so $\e_{0\cdots 9}=-1$).

The matrix $\mathcal{C}_{AB}$ is an antisymmetric charge conjugation matrix which we use for raising and lowering spinor indices.  It satisfies the useful identities
\be
\label{eq:ChargeTranspose}
\mathcal{C}\G^\m\mathcal{C}^{-1}=-\lp\G^\m\rp^T,\qquad\mathcal{C}\G_{11}\mathcal{C}^{-1}=-\lp\G_{11}\rp^T.
\ee

\section{Computation of $(-1)^F$ on the boundary state}
\label{sec:BoundaryStateFermionNumber}

The operator $(-1)^F$ commutes with everything outside of the $\psi$ and $\phi$ sectors, so we shall ignore those other sectors (which are also independent of $\eta$).  Then we will use the correlators of section \ref{subsec:Correlators} to argue that (\ref{eq:NSBoundaryStateFermion}) and (\ref{eq:RBoundaryStateFermion}) hold.  We won't work through the complete details, but rather sketch how this can be done.

In the NS sector, we use the fact that $(-1)^F$ acts as $-1$ on the $-1$-picture vacuum, and anticommutes with left moving fermions, to establish for example
\be
\left\langle -1,-1\left|\lp -1\rp^F\right|B;\eta\right\rangle_{NS}=-1=-\left\langle -1,-1|B;-\eta\right\rangle_{NS},
\ee
and
\begin{multline}
\left\langle -1,-1\left|\psi^\m(z_1)\wtpsi^\n(\barz_2)\lp -1\rp^F\right|B;\eta\right\rangle_{NS}=\frac{-i\eta D^{\m\n}}{z_1\barz_2-1} \\
=-\left\langle -1,-1\left|\psi^\m(z_1)\wtpsi^\n(barz_2)\right|B;-\eta\right\rangle_{NS},
\end{multline}
It is not difficult to show that correlators with arbitrary many $\psi$, $\wtpsi$, $\phi$, and $\wtphi$ insertions will obey similar expressions and thus that
\be
\lp -1\rp^F\left|B;\eta\right\rangle_{NS}=-\left|B;-\eta\right\rangle_{NS}.
\ee
The right-moving fermion number works exactly the same way, and one finds also
\be
\lp -1\rp^{\widetilde{F}}\left|B;\eta\right\rangle_{NS}=-\left|B;-\eta\right\rangle_{NS}.
\ee

In the R sector, we will make use of (\ref{eq:RamondStateFermionNumber}) to proceed in a similar fashion,
\bea
\label{eq:RGSOExamples}
\left\langle -\hlf,-\frac{3}{2};A,B\left|\lp -1\rp^F\right|B;\eta\right\rangle_R &=& -i\eta\ls\G_{11}\mathcal{C}^{-1}\mathcal{M}(\eta)\mathcal{C}^{-1}\rs^{AB},\non\\
\left\langle -\frac{3}{2},-\hlf;A,B\left|\lp -1\rp^F\right|B;\eta\right\rangle_R &=& -\ls\G_{11}\mathcal{C}^{-1}\mathcal{M}(\eta)\mathcal{C}^{-1}\rs^{AB},\\
\left\langle -\hlf,-\frac{3}{2};A,B\left|\lp -1\rp^{\widetilde{F}}\right|B;\eta\right\rangle_R &=& i\eta\ls\mathcal{C}^{-1}\mathcal{M}(\eta)\mathcal{C}^{-1}\G_{11}^T\rs^{AB},\non\\
\left\langle -\frac{3}{2},-\hlf;A,B\left|\lp -1\rp^{\widetilde{F}}\right|B;\eta\right\rangle_R &=& \ls\mathcal{C}^{-1}\mathcal{M}(\eta)\mathcal{C}^{-1}\G_{11}^T\rs^{AB}.\non
\eea
Using (\ref{eq:MExpression}), we have
\be
\G_{11}\mathcal{C}^{-1}\mathcal{M}(\eta)=-\mathcal{C}^{-1}\mathcal{M}(-\eta),\qquad\mathcal{M}(\eta)\mathcal{C}^{-1}\G_{11}^T=\lp -1\rp^{p+1}\mathcal{M}(-\eta)\mathcal{C}^{-1},
\ee
which, comparing with
\bea
\left\langle -\hlf,-\frac{3}{2};A,B|B;\eta\right\rangle_R &=& -i\eta\ls\mathcal{C}^{-1}\mathcal{M}(\eta)\mathcal{C}^{-1}\rs^{AB},\\
\left\langle -\frac{3}{2},-\hlf;A,B|B;\eta\right\rangle_R &=& \ls\mathcal{C}^{-1}\mathcal{M}(\eta)\mathcal{C}^{-1}\rs^{AB},
\eea
implies that all of (\ref{eq:RGSOExamples}) are consistent with
\be
\label{eq:RBFN}
\lp -1\rp^F\left|B;\eta\right\rangle_R=\left|B;-\eta\right\rangle_R,\qquad\lp -1\rp^{\widetilde{F}}\left|B;\eta\right\rangle_R=\lp -1\rp^{p+1}\left|B;-\eta\right\rangle_R.
\ee

One can show that similar expressions hold for all correlators, which establishes (\ref{eq:RBFN}).

\section{Details of field theory computations}

\subsection{Two NS-NS fields}
\label{subsec:NSNSAppendix}

In this section we will do the detailed field theory computations for the amplitudes involving two NS-NS fields interacting with a type II D$p$-brane.

We start with the interaction between a dilaton $\Phi$ and a graviton $h_{\m\n}$.
\begin{multline}
\label{eq:PhihFieldTheory}
\left\langle\Phi h\right\rangle\sim\Phi h_{\m\n}\left\{\lp\frac{\d}{\d\Phi}\frac{\d}{\d h_{\m\n}}S_{p+1}\rp+i\lp\frac{\d}{\d\Phi}\frac{\d}{\d h_{\m\n}}\frac{\d}{\d\Phi}S_{10}\rp G^{(\Phi)}\lp\frac{\d}{\d\Phi}S_{p+1}\rp\right.\\
\left.+i\lp\frac{\d}{\d\Phi}\frac{\d}{\d X^i}S_{p+1}\rp G^{(X)\,ij}\lp\frac{\d}{\d h_{\m\n}}\frac{\d}{\d X^j}S_{p+1}\rp\right\}.
\end{multline}
In order to use this formula we need to compute some variations of the bulk and brane actions (note that we already computed $\d_\Phi\d_XS_{p+1}$ in section \ref{subsec:NSNSSUGRA}),
\bea
h_{\m\n}\frac{\d}{\d\Phi}\frac{\d}{\d h_{\m\n}}S_{p+1} &=& -\m_p\k^2\frac{p-3}{2\sqrt{2}}h^a_{\hphantom{a}a},\non\\
h_{\m\n}\frac{\d}{\d\Phi}\frac{\d}{\d h_{\m\n}}\frac{\d}{\d\Phi}S_{10} &=& 2\k h^{\m\n}p_{1\,\m}p_{1\,\n},\non\\
\frac{\d}{\d\Phi}S_{p+1} &=& -\m_p\k\frac{p-3}{2\sqrt{2}},\\
h_{\m\n}\frac{\d}{\d h_{\m\n}}\frac{\d}{\d X^i}S_{p+1} &=& i\sqrt{\m_p}\k\lp -h^a_{\hphantom{a}a}p_{2\,i}+2h^a_{\hphantom{a}i}p_{2\,a}\rp,\non
\eea
It is important to emphasize that we assume in these expressions that external states (though not the propagating internal lines of course) are on-shell, so we drop terms such as $h^\m_{\hphantom{\m}\m}$ or $h^{\m\n}p_{2\,\n}$.  Plugging these results into (\ref{eq:PhihFieldTheory}), we get
\begin{multline}
\Phi\left\{-\m_p\k^2\frac{p-3}{2\sqrt{2}}h^a_{\hphantom{a}a}+ih_{\m\n}\lp-\k t\eta^{\m\n}+\k\lp 2p_1^\m p_1^\n+ p_1^\m p_2^\n+p_2^\m p_1^\n\rp\rp\frac{-i}{2t}\lp -\m_p\k\frac{p-3}{2\sqrt{2}}\rp\right.\\
\left.+i\lp-i\sqrt{\m_p}\k\frac{p-3}{2\sqrt{2}}p_{1\,i}\rp\frac{-i\d^{ij}}{s/2}\lp-i\sqrt{\m_p}\k p_{2\,j}h^a_{\hphantom{a}a}+2i\sqrt{\m_p}\k p_2^ah_{aj}\rp\right\}\\
=-\m_p\k^2\frac{p-3}{2\sqrt{2}}\Phi\left\{\frac{1}{t}h_{\m\n}p_1^\m p_1^\n+\frac{2t}{s}h^a_{\hphantom{a}a}+\frac{4}{s}h_{ai}p_1^ap_1^i+2h^a_{\hphantom{a}a}\right\},
\end{multline}
in precise agreement with (\ref{eq:NSFieldTwoPoint}), including the same normalization constant as in the $\langle\Phi\Phi\rangle$ amplitude of section \ref{subsec:NSNSSUGRA}.

Next we turn to the interaction of two gravitons,
\begin{multline}
\left\langle hh\right\rangle\sim h_{1\,\m\n}h_{2\,\rho\s}\left\{\lp\frac{\d}{\d h_{1\,\m\n}}\frac{\d}{\d h_{2\,\rho\s}}S_{p+1}\rp\right.\\
\left.+i\lp\frac{\d}{\d h_{1\,\m\n}}\frac{\d}{\d h_{2\,\rho\s}}\frac{\d}{\d h_{\tau\lambda}}S_{10}\rp G^{(h)}_{\tau\lambda,\om\varphi}\lp\frac{\d}{\d h_{\om\varphi}}S_{p+1}\rp\right.\\
\left.+i\lp\frac{\d}{\d h_{1\,\m\n}}\frac{\d}{\d X^i}S_{p+1}\rp G^{(X)\,ij}\lp\frac{\d}{\d h_{2\,\rho\s}}\frac{\d}{\d X^j}S_{p+1}\rp\right\}.
\end{multline}
The additional variations we will need are
\be
h_{1\,\m\n}h_{2\,\rho\s}\frac{\d}{\d h_{1\,\m\n}}\frac{\d}{\d h_{2\,\rho\s}}S_{p+1}=\m_p\k^2\lp -h_{1\,a}^ah_{2\,b}^b+2h_1^{ab}h_{2\,ab}\rp,
\ee
and the cubic graviton interaction from the Einstein-Hilbert term in the bulk,
\begin{multline}
h_{1\,\rho\s}h_{2\,\tau\lambda}\frac{\d}{\d h_{1\,\rho\s}}\frac{\d}{\d h_{2\,\tau\lambda}}\frac{\d}{\d h_{\m\n}}S_{10}=\k\ls -3th_1^{\rho\s}h_{2\,\rho\s}\eta^{\m\n}+2h_1^{\rho\s}h_{2\,\rho}^{\hphantom{2\,\rho}\tau}p_{1\,\tau}p_{2\,\s}\eta^{\m\n}\right.\\
\left.+4th_1^{(\m|\rho|}h_{2\,\rho}^{\n)}+2h_1^{\m\n}h_2^{\rho\s}p_{1\,\rho}p_{1\,\s}+2h_1^{\rho\s}h_2^{\m\n}p_{2\,\rho}p_{2\,\s}-4h_1^{(\m|\rho|}h_2^{\n)\s}p_{1\,\s}p_{2\,\rho}\right.\\
\left.-4h_1^{(\m|\rho}h_{2\,\rho}^{\hphantom{2\,\rho}\s|}p_1^{\n)}p_{1\,\s}-4h_1^{\rho\s}h_{2\,\rho}^{(\m}p_2^{\n)}p_{2\,\s}+2h_1^{\rho\s}h_{2\,\rho\s}\lp p_1^\m p_1^\n+p_2^\m p_2^\n\rp+2h_1^{\rho\s}h_{2\,\rho\s}p_1^{(\m}p_2^{\n)}\rs,\non
\end{multline}
where we again emphasize that we assume that the external gravitons $h_1$ and $h_2$ are on-shell.  Then $\langle hh\rangle$ becomes
\begin{multline}
\m_p\k^2\lp -h_{1\,a}^ah_{2\,b}^b+2h_1^{ab}h_{2\,ab}\rp+\m_p\k^2\frac{p-7}{16t}\ls -3t\lp p+1\rp h_1^{\m\n}h_{2\,\m\n}\right.\\
\left.+2\lp p+1\rp h_1^{\m\n}h_{2\,\m}^{\hphantom{2\,\m}\rho}p_{1\,\rho}p_{2\,\n}+4th_1^{a\m}h_{2\,a\m}+2h_{1\,a}^ah_2^{\m\n}p_{1\,\m}p_{1\,\n}+2h_1^{\m\n}h_{2\,a}^ap_{2\,\m}p_{2\,\n}\right.\\
\left.-4h_1^{a\m}h_{2\,a}^{\hphantom{2\,a}\n}p_{1\,\n}p_{2\,\m}+4h_1^{a\m}h_{2\,\m}^{\hphantom{2\,\m}\n}p_{1\,\n}p_{2\,a}+4h_1^{\m\n}h_{2\,\m}^ap_{1\,a}p_{2\,\n}+sh_1^{\m\n}h_{2\,\m\n}\rs\\
+\m_p\k^2\frac{p+1}{16t}\ls-3t\lp 9-p\rp h_1^{\m\n}h_{2\,\m\n}+2\lp 9-p\rp h_1^{\m\n}h_{2\,\m}^{\hphantom{2\,\m}\rho}p_{1\,\rho}p_{2\,\n}+4th_1^{i\m}h_{2\,i\m}\right.\\
\left.-2h_{1\,a}^ah_2^{\m\n}p_{1\,\m}p_{1\,\n}-2h_1^{\m\n}h_{2\,a}^ap_{2\,\m}p_{2\,\n}-4h_1^{i\m}h_{2\,i}^{\hphantom{2\,i}\n}p_{1\,\n}p_{2\,\m}-4h_1^{a\m}h_{2\,\m}^{\hphantom{2\,\m}\n}p_{1\,\n}p_{2\,a}\right.\\
\left.-4h_1^{\m\n}h_{2\,\m}^ap_{1\,a}p_{2\,\n}+\lp 2t-s\rp h_1^{\m\n}h_{2\,\m\n}\rs\\
-\m_p\k^2\frac{2}{s}\ls\lp t+\frac{s}{2}\rp h_{1\,a}^ah_{2\,b}^b+2h_{1\,a}^ah_2^{bi}p_{1\,b}p_{1\,i}+2h_1^{ai}h_{2\,b}^bp_{2\,a}p_{2\,i}+4h_1^{ai}h_{2\,i}^bp_{1\,b}p_{2\,a}\rs\\
=-\m_p\k^2\left\{\frac{1}{t}\ls\frac{s}{2}h_1^{\m\n}h_{2\,\m\n}+h_{1\,a}^ah_2^{\m\n}p_{1\,\m}p_{1\,\n}+h_1^{\m\n}h_{2\,a}^ap_{2\,\m}p_{2\,\n}+2h_1^{ab}h_{2\,a}^{\hphantom{2\,a}c}p_{1\,c}p_{2\,b}\right.\right.\\
\left.\left.-2h_1^{ai}h_{2\,a}^{\hphantom{2\,a}j}p_{1\,j}p_{2\,i}+4h_1^{ai}h_{2\,i}^{\hphantom{2\,i}b}p_{1\,b}p_{2\,a}+2h_1^{ai}h_{2\,i}^{\hphantom{2\,i}j}p_{1\,j}p_{2\,a}+2h_1^{ij}h_{2\,i}^ap_{1\,a}p_{2\,j}\rs\right.\\
\left.+\frac{1}{s}\ls 2th_{1\,a}^ah_{2\,b}^b+4h_{1\,a}^ah_2^{bi}p_{1\,b}p_{1\,i}+4h_1^{ai}h_{2\,b}^bp_{2\,a}p_{2\,i}+8h_1^{ai}h_{2\,i}^bp_{1\,b}p_{2\,a}\rs\right.\\
\left.+2h_{1\,a}^ah_{2\,b}^b+2h_1^{ai}h_{2\,ai}\right\},
\end{multline}
once again in precise agreement with (\ref{eq:NSFieldTwoPoint}).

Finally we turn to the interaction of two $B$-fields with the brane,
\begin{multline}
\left\langle BB\right\rangle\sim B_{1\,\m\n}B_{2\,\rho\s}\left\{\lp\frac{\d}{\d B_{1\,\m\n}}\frac{\d}{\d B_{2\,\rho\s}}S_{p+1}\rp+i\lp\frac{\d}{\d B_{1\,\m\n}}\frac{\d}{\d B_{2\,\rho\s}}\frac{\d}{\d\Phi}S_{10}\rp G^{(\Phi)}\lp\frac{\d}{\d\Phi}S_{p+1}\rp\right.\\
\left.+i\lp\frac{\d}{\d B_{1\,\m\n}}\frac{\d}{\d B_{2\,\rho\s}}\frac{\d}{\d h_{\tau\lambda}}S_{10}\rp G^{(h)}_{\tau\lambda,\om\varphi}\lp\frac{\d}{\d h_{\om\varphi}}S_{p+1}\rp\right.\\
\left.+i\lp\frac{\d}{\d B_{1\,\m\n}}\frac{\d}{\d A_a}S_{p+1}\rp G^{(A)}_{ab}\lp\frac{\d}{\d B_{2\,\rho\s}}\frac{\d}{\d A_b}S_{p+1}\rp\right\}.
\end{multline}
We need vertices on the brane
\be
B_{1\,\m\n}B_{2\,\rho\s}\frac{\d}{\d B_{1\,\m\n}}\frac{\d}{\d B_{2\,\rho\s}}S_{p+1}=-2\m_p\k^2B_1^{ab}B_{2\,ab},
\ee
and
\be
B_{1\,\m\n}\frac{\d}{\d B_{1\,\m\n}}\frac{\d}{\d A_a}S_{p+1}=2i\sqrt{\m_p}\k B_1^{ab}p_{2\,b},
\ee
and bulk interactions
\be
B_{1\,\m\n}B_{2\,\rho\s}\frac{\d}{\d B_{1\,\m\n}}\frac{\d}{\d B_{2\,\rho\s}}\frac{\d}{\d\Phi}S_{10}=-\sqrt{2}\k tB_1^{\m\n}B_{2\,\m\n}+2\sqrt{2}\k B_1^{\m\n}B_{2\,\m}^{\hphantom{2\,\m}\rho}p_{1\,\rho}p_{2\,\n},
\ee
and
\begin{multline}
B_{1\,\rho\s}B_{2\,\tau\lambda}\frac{\d}{\d B_{1\,\rho\s}}\frac{\d}{\d B_{2\,\tau\lambda}}\frac{\d}{\d h_{\m\n}}S_{10}=\k\ls tB_1^{\rho\s}B_{2\,\rho\s}\eta^{\m\n}-2B_1^{\rho\s}B_{2\,\rho}^{\hphantom{2\,\rho}\tau}p_{1\,\tau}p_{2\,\s}\eta^{\m\n}\right.\\
\left.-2B_1^{\rho\s}B_{2\,\rho\s}p_1^{(\m}p_2^{\n)}-4tB_1^{(\m|\rho|}B_{2\,\rho}^{\n)}+4B_1^{(\m|\rho|}B_2^{\n)\s}p_{1\,\s}p_{2\,\rho}\right.\\
\left.-4B_1^{(\m|\rho}B_{2\,\rho}^{\hphantom{2\,\rho}\s|}p_{1\,\s}p_2^{\n)}-4B_1^{\rho\s}B_{2\,\rho}^{(\m}p_1^{\n)}p_{2\,\s}\rs.
\end{multline}
These then lead to an amplitude
\begin{multline}
-2\m_p\k^2B_1^{ab}B_{2\,ab}+\m_p\k^2\frac{p-3}{4\sqrt{2}t}\ls\sqrt{2} tB_1^{\m\n}B_{2\,\m\n}-2\sqrt{2}B_1^{\m\n}B_{2\,\m}^{\hphantom{2\,\m}\rho}p_{1\,\rho}p_{2\,\n}\rs\\
+\m_p\k^2\frac{p-7}{16t}\ls\lp p+1\rp tB_1^{\m\n}B_{2\,\m\n}-2\lp p+1\rp B_1^{\m\n}B_{2\,\m}^{\hphantom{2\,\m}\rho}p_{1\,\rho}p_{2\,\n}+sB_1^{\m\n}B_{2\,\m\n}\right.\\
\left.-4tB_1^{a\m}B_{2\,a\m}+4B_1^{a\m}B_{2\,a}^{\hphantom{2\,a}\n}p_{1\,\n}p_{2\,\m}-4B_1^{a\m}B_{2\,\m}^{\hphantom{2\,\m}\n}p_{1\,\n}p_{2\,a}-4B_1^{\m\n}B_{2\,\m}^ap_{1\,a}p_{2\,\n}\rs\\
+\m_p\k^2\frac{p+1}{16t}\ls\lp 9-p\rp tB_1^{\m\n}B_{2\,\m\n}-2\lp 9-p\rp B_1^{\m\n}B_{2\,\m}^{\hphantom{2\,\m}\rho}p_{1\,\rho}p_{2\,\n}-\lp 2t+s\rp B_1^{\m\n}B_{2\,\m\n}\right.\\
\left.-4tB_1^{i\m}B_{2\,i\m}+4B_1^{i\m}B_{2\,i}^{\hphantom{2\,i}\n}p_{1\,\n}p_{2\,\m}-4B_1^{i\m}B_{2\,\m}^{\hphantom{2\,\m}\n}p_{1\,\n}p_{2\,i}-4B_1^{\m\n}B_{2\,\m}^ip_{1\,i}p_{2\,\n}\rs\\
-\m_p\k^2\frac{8}{s}B_1^{ab}B_{2\,a}^{\hphantom{2\,a}c}p_{1\,c}p_{2\,b}\\
=-\m_p\k^2\left\{\frac{1}{t}\ls\frac{s}{2}B_1^{\m\n}B_{2\,\m\n}+4B_1^{ab}B_{2\,a}^{\hphantom{2\,a}c}p_{1\,c}p_{2\,b}+2B_1^{ab}B_{2\,a}^{\hphantom{2\,a}i}p_{1\,i}p_{2\,b}+2B_1^{ai}B_{2\,a}^{\hphantom{2\,a}b}p_{1\,b}p_{2\,i}\right.\right.\\
\left.\left.+2B_1^{ai}B_{2\,i}^bp_{1\,b}p_{2\,a}-2B_1^{ij}B_{2\,i}^{\hphantom{2\,i}k}p_{1\,k}p_{2\,j}\vphantom{\frac{s}{2}}\rs\right.\\
\left.+\frac{8}{s}B_1^{ab}B_{2\,a}^{\hphantom{2\,a}c}p_{1\,c}p_{2\,b}+B_1^{ab}B_{2\,ab}+B_1^{ij}B_{2\,ij}\right\}.
\end{multline}
Comparing with (\ref{eq:NSFieldTwoPoint}), it is gratifying to note that all the two-point functions agree.

\subsection{One NS-NS and one R-R field}
\label{subsec:RRAppendix}

Here we give the details for the field theory computations of two-point functions involving one R-R potential and one NS-NS field.  The first one is
\begin{multline}
\left\langle C^{(p+3)}B\right\rangle\sim iC^{(p+3)}_{\m_1\cdots\m_{p+3}}B_{\n\rho}\lp\frac{\d}{\d C^{(p+3)}_{\m_1\cdots\m_{p+3}}}\frac{\d}{\d B_{\n\rho}}\frac{\d}{\d C^{(p+1)}_{\s_1\cdots\s_{p+1}}}S_{10}\rp\\
\times G^{(C^{(p+1)})}_{\s_1\cdots\s_{p+1},\tau_1\cdots\tau_{p+1}}\lp\frac{\d}{\d C^{(p+1)}_{\tau_1\cdots\tau_{p+1}}}S_{p+1}\rp.
\end{multline}
Since only one diagram contributes, the only variations we need to compute are the one-point contact term
\be
\frac{\d}{\d C^{(p+1)}_{a_1\cdots a_{p+1}}}S_{p+1}=\frac{\sqrt{2}\m_p\k}{(p+1)!}\e^{a_1\cdots a_{p+1}},
\ee
and the bulk interaction
\begin{multline}
C^{(p+3)}_{\n_1\cdots\n_{p+3}}B_{\rho\s}\frac{\d}{\d C^{(p+3)}_{\n_1\cdots\n_{p+3}}}\frac{\d}{\d B_{\rho\s}}\frac{\d}{\d C^{(p+1)}_{\m_1\cdots\m_{p+1}}}S_{10}\\
=\k\ls\frac{1}{(p+1)!}C^{{\rlap{$\mathsurround=0pt\scriptstyle{(p+3)}$}}\hphantom{\m_1\cdots\m_{p+1}}\n\rho}_{\m_1\cdots\m_{p+1}}\lp tB_{\n\rho}-2B_\n^{\hphantom{\n}\s}p_{1\,\s}p_{2\,\rho}\rp-\frac{1}{p!}C^{{\rlap{$\mathsurround=0pt\scriptstyle{(p+3)}$}}\hphantom{[\m_1\cdots\m_p}\n\rho\s}_{[\m_1\cdots\m_p}B_{|\n\rho|}p_{1\,\m_{p+1}]}p_{2\,\s}\rs.
\end{multline}
We can now evaluate the amplitude $\langle C^{(p+3)}B\rangle$ as
\begin{multline}
\frac{\m_p\k^2}{\sqrt{2}t}\e^{a_1\cdots a_{p+1}}\left\{\frac{1}{(p+1)!}C^{{\rlap{$\mathsurround=0pt\scriptstyle{(p+3)}$}}\hphantom{a_1\cdots a_{p+1}}ij}_{a_1\cdots a_{p+1}}\lp tB_{ij}+2B^b_{\hphantom{b}i}p_{1\,b}p_{2\,j}-2B_i^{\hphantom{i}k}p_{1\,k}p_{2\,j}\rp\right.\\
\left.-\frac{1}{p!}\ls C^{{\rlap{$\mathsurround=0pt\scriptstyle{(p+3)}$}}\hphantom{a_1\cdots a_p}bij}_{a_1\cdots a_p}\lp 2B_{bi}p_{1\,a_{p+1}}p_{2\,j}+B_{ij}p_{1\,a_{p+1}}p_{2\,b}\rp+C^{{\rlap{$\mathsurround=0pt\scriptstyle{(p+3)}$}}\hphantom{a_1\cdots a_p}ijk}_{a_1\cdots a_p}B_{ij}p_{1\,a_{p+1}}p_{2\,k}\rs\right\}\\
=\frac{\m_p\k^2}{\sqrt{2}t}\e^{a_1\cdots a_{p+1}}\left\{\frac{1}{(p+1)!}C^{{\rlap{$\mathsurround=0pt\scriptstyle{(p+3)}$}}\hphantom{a_1\cdots a_{p+1}}ij}_{a_1\cdots a_{p+1}}\lp rB_{ij}-2B_i^{\hphantom{i}k}p_{1\,k}p_{2\,j}\rp\right.\\
\left.-\frac{1}{p!}C^{{\rlap{$\mathsurround=0pt\scriptstyle{(p+3)}$}}\hphantom{a_1\cdots a_p}ijk}_{a_1\cdots a_p}B_{ij}p_{1\,a_{p+1}}p_{2\,k}\right\}.
\end{multline}
Recall that we have defined $r=t+\frac{s}{2}$.  This expression can be rewritten as something proportional to $\frac{1}{t}F^{{\rlap{$\mathsurround=0pt\scriptstyle{(p+4)}$}}\hphantom{a_1\cdots a_{p+1}}ijk}_{a_1\cdots a_{p+1}}H_{ijk}$ in agreement with (\ref{eq:F(p+4)Coupling}).

Next we turn to the $\langle C^{(p-1)}B\rangle$ amplitude, which is the only one in this section which receives contributions from three different diagrams,
\begin{multline}
\left\langle C^{(p-1)}B\right\rangle\sim C^{(p-1)}_{\m_1\cdots\m_{p-1}}B_{\n\rho}\left\{\lp\frac{\d}{\d C^{(p-1)}_{\m_1\cdots\m_{p-1}}}\frac{\d}{\d B_{\n\rho}}S_{p+1}\rp\right.\\
\left.+i\lp\frac{\d}{\d C^{(p-1)}_{\m_1\cdots\m_{p-1}}}\frac{\d}{\d B_{\n\rho}}\frac{\d}{\d C^{(p+1)}_{\s_1\cdots\s_{p+1}}}S_{10}\rp G^{(C^{(p+1)})}_{\s_1\cdots\s_{p+1},\tau_1\cdots\tau_{p+1}}\lp\frac{\d}{\d C^{(p+1)}_{\tau_1\cdots\tau_{p+1}}}S_{p+1}\rp\right.\\
\left.+i\lp\frac{\d}{\d C^{(p-1)}_{\m_1\cdots\m_{p-1}}}\frac{\d}{\d A_a}S_{p+1}\rp G^{(A)}_{ab}\lp\frac{\d}{\d B_{\n\rho}}\frac{\d}{\d A_b}S_{p+1}\rp\right\}.
\end{multline}
As usual, we need to compute some contact terms,
\be
C^{(p-1)}_{\m_1\cdots\m_{p-1}}B_{\n\rho}\frac{\d}{\d C^{(p-1)}_{\m_1\cdots\m_{p-1}}}\frac{\d}{\d B_{\n\rho}}S_{p+1}=\frac{\sqrt{2}\m_p\k^2}{(p-1)!}\e^{a_1\cdots a_{p+1}}C^{(p-1)}_{a_1\cdots a_{p-1}}B_{a_pa_{p+1}},
\ee
\be
C^{(p-1)}_{\m_1\cdots\m_{p-1}}\frac{\d}{\d C^{(p-1)}_{\m_1\cdots\m_{p-1}}}\frac{\d}{\d A_a}S_{p+1}=i\frac{\sqrt{2\m_p}\k}{(p-1)!}\e^{b_1\cdots b_pa}C^{(p-1)}_{b_1\cdots b_{p-1}}p_{2\,b_p},
\ee
as well as a bulk interaction
\begin{multline}
C^{(p-1)}_{\n_1\cdots\n_{p-1}}B_{\rho\s}\frac{\d}{\d C^{(p-1)}_{\n_1\cdots\n_{p-1}}}\frac{\d}{\d B_{\rho\s}}\frac{\d}{\d C^{(p+1)}_{\m_1\cdots\m_{p+1}}}S_{10}\\
=\k\ls\frac{1}{(p-1)!}C^{(p-1)}_{\m_1\cdots\m_{p-1}}\lp -tB_{\m_p\m_{p+1}}+2B_{\m_p}^{\hphantom{\m_p}\n}p_{1\,|\n|}p_{2\,\m_{p+1}}\rp\right.\\
\left.+\frac{1}{(p-2)!}C^{{\rlap{$\mathsurround=0pt\scriptstyle{(p-1)}$}}\hphantom{[\m_1\cdots\m_{p-2}}\n}_{[\m_1\cdots\m_{p-2}}B_{\m_{p-1}\m_p}p_{2\,\m_{p+1}]}p_{2\,\n}\rs.
\end{multline}
Plugging into the amplitude, we find
\begin{multline}
\sqrt{2}\m_p\k^2\e^{a_1\cdots a_{p+1}}\left\{\frac{1}{(p-1)!}C^{(p-1)}_{a_1\cdots a_{p-1}}B_{a_pa_{p+1}}\right.\\
\left.+\frac{1}{2t}\ls\frac{1}{(p-1)!}C^{(p-1)}_{a_1\cdots a_{p-1}}\lp -tB_{a_pa_{p+1}}+2B_{a_p}^{\hphantom{a_p}b}p_{1\,b}p_{2\,a_{p+1}}+2B_{a_p}^{\hphantom{a_p}i}p_{1\,i}p_{2\,a_{p+1}}\rp\right.\right.\\
\left.\left.+\frac{1}{(p-2)!}\lp C^{{\rlap{$\mathsurround=0pt\scriptstyle{(p-1)}$}}\hphantom{a_1\cdots a_{p-2}}b}_{a_1\cdots a_{p-2}}B_{a_{p-1}a_p}p_{2\,a_{p+1}}p_{2\,b}+C^{{\rlap{$\mathsurround=0pt\scriptstyle{(p-1)}$}}\hphantom{a_1\cdots a_{p-2}}i}_{a_1\cdots a_{p-2}}B_{a_{p-1}a_p}p_{2\,a_{p+1}}p_{2\,i}\rp\rs\right.\\
\left.+\frac{4}{s}\frac{1}{(p-1)!}C^{(p-1)}_{a_1\cdots a_{p-1}}B_{a_p}^{\hphantom{a_p}b}p_{1\,b}p_{2\,a_{p+1}}\right\}\\
=\sqrt{2}\m_p\k^2\e^{a_1\cdots a_{p+1}}\left\{\frac{1}{(p-1)!}C^{(p-1)}_{a_1\cdots a_{p-1}}\lp\frac{r}{2t}B_{a_pa_{p+1}}-\frac{4r}{st}B_{a_p}^{\hphantom{a_p}b}p_{1\,a_{p+1}}p_{1\,b}-\frac{1}{t}B_{a_p}^{\hphantom{a_p}i}p_{1\,a_{p+1}}p_{1\,i}\rp\right.\\
\left.-\frac{1}{(p-2)!}C^{{\rlap{$\mathsurround=0pt\scriptstyle{(p-1)}$}}\hphantom{a_1\cdots a_{p-2}}i}_{a_1\cdots a_{p-2}}B_{a_{p-1}a_p}p_{1\,a_{p+1}}p_{2\,i}\right\}.
\end{multline}
This result agrees with the amplitude (\ref{eq:(-1,0)C(p-1)}).

Finally, we have the couplings to $C^{(p+1)}$.  In each case there is no direct contact term but there are two contributing diagrams, in the $t$ channel and $s$ channel respectively.  For the dilaton we have
\begin{multline}
\left\langle C^{(p+1)}\Phi\right\rangle\sim C^{(p+1)}_{\m_1\cdots\m_{p+1}}\Phi\left\{ i\lp\frac{\d}{\d C^{(p+1)}_{\m_1\cdots\m_{p+1}}}\frac{\d}{\d\Phi}\frac{\d}{\d C^{(p+1)}_{\n_1\cdots\n_{p+1}}}S_{10}\rp\right.\\
\left.\times G^{(C^{(p+1)})}_{\n_1\cdots\n_{p+1},\rho_1\cdots\rho_{p+1}}\lp\frac{\d}{\d C^{(p+1)}_{\rho_1\cdots\rho_{p+1}}}S_{p+1}\rp\right.\\
\left.+i\lp\frac{\d}{\d C^{(p+1)}_{\m_1\cdots\m_{p+1}}}\frac{\d}{\d X^i}S_{p+1}\rp G^{(X)\,ij}\lp\frac{\d}{\d\Phi}\frac{\d}{\d X^j}S_{p+1}\rp\right\}.
\end{multline}
The requisite variations are
\begin{multline}
C^{(p+1)}_{\m_1\cdots\m_{p+1}}\frac{\d}{\d C_{\m_1\cdots\m_{p+1}}}\frac{\d}{\d X^i}S_{p+1}\\
=i\sqrt{2\m_p}\k\e^{a_1\cdots a_{p+1}}\lp\frac{1}{(p+1)!}C^{(p+1)}_{a_1\cdots a_{p+1}}p_{1\,i}-\frac{1}{p!}C^{(p+1)}_{a_1\cdots a_pi}p_{1\,a_{p+1}}\rp,
\end{multline}
and
\begin{multline}
C^{(p+1)}_{\n_1\cdots\n_{p+1}}\frac{\d}{\d C^{(p+1)}_{\n_1\cdots\n_{p+1}}}\frac{\d}{\d\Phi}\frac{\d}{\d C^{(p+1)}_{\m_1\cdots\m_{p+1}}}S_{10}\\
=\frac{p-3}{\sqrt{2}}\k\ls\frac{1}{(p+1)!}tC^{(p+1)}_{\m_1\cdots\m_{p+1}}-\frac{1}{p!}C^{{\rlap{$\mathsurround=0pt\scriptstyle{(p+1)}$}}\hphantom{[\m_1\cdots\m_p}\n}_{[\m_1\cdots\m_p}p_{1\,\m_{p+1}]}p_{2\,\n}\rs.
\end{multline}
These lead to an amplitude
\begin{multline}
\lp p-3\rp\m_p\k^2\Phi\e^{a_1\cdots a_{p+1}}\left\{\frac{1}{2t}\ls\frac{1}{(p+1)!}tC^{(p+1)}_{a_1\cdots a_{p+1}}-\frac{1}{p!}C^{{\rlap{$\mathsurround=0pt\scriptstyle{(p+1)}$}}\hphantom{a_1\cdots a_p}b}_{a_1\cdots a_p}p_{1\,a_{p+1}}p_{2\,b}\right.\right.\\
\left.\left.-\frac{1}{p!}C^{{\rlap{$\mathsurround=0pt\scriptstyle{(p+1)}$}}\hphantom{a_1\cdots a_p}i}_{a_1\cdots a_p}p_{1\,a_{p+1}}p_{2\,i}\rs+\frac{1}{s}\ls\frac{1}{(p+1)!}rC^{(p+1)}_{a_1\cdots a_{p+1}}-\frac{1}{p!}C^{{\rlap{$\mathsurround=0pt\scriptstyle{(p+1)}$}}\hphantom{a_1\cdots a_p}i}_{a_1\cdots a_p}p_{1\,a_{p+1}}p_{2\,i}\rs\right\}\\
=\lp p-3\rp\m_p\k^2\Phi\e^{a_1\cdots a_{p+1}}\left\{\frac{1}{(p+1)!}\frac{r^2}{st}C^{(p+1)}_{a_1\cdots a_{p+1}}-\frac{1}{p!}\frac{r}{st}C^{{\rlap{$\mathsurround=0pt\scriptstyle{(p+1)}$}}\hphantom{a_1\cdots a_p}i}_{a_1\cdots a_p}p_{1\,a_{p+1}}p_{2\,i}\right\}.
\end{multline}

And for the coupling of $C^{(p+1)}$ to a graviton we have
\begin{multline}
\left\langle C^{(p+1)}h\right\rangle\sim C^{(p+1)}_{\m_1\cdots\m_{p+1}}h_{\n\rho}\left\{ i\lp\frac{\d}{\d C^{(p+1)}_{\m_1\cdots\m_{p+1}}}\frac{\d}{\d h_{\n\rho}}\frac{\d}{\d C^{(p+1)}_{\s_1\cdots\s_{p+1}}}S_{10}\rp\right.\\
\left.\times G^{(C^{(p+1)})}_{\s_1\cdots\s_{p+1},\tau_1\cdots\tau_{p+1}}\lp\frac{\d}{\d C^{(p+1)}_{\tau_1\cdots\tau_{p+1}}}S_{p+1}\rp\right.\\
\left.+i\lp\frac{\d}{\d C^{(p+1)}_{\m_1\cdots\m_{p+1}}}\frac{\d}{\d X^i}S_{p+1}\rp G^{(X)\,ij}\lp\frac{\d}{\d h_{\n\rho}}\frac{\d}{\d X^j}S_{p+1}\rp\right\}.
\end{multline}
The only variation we're missing is
\begin{multline}
C^{(p+1)}_{\n_1\cdots\n_{p+1}}h_{\rho\s}\frac{\d}{\d C^{(p+1)}_{\n_1\cdots\n_{p+1}}}\frac{\d}{\d h_{\rho\s}}\frac{\d}{\d C^{(p+1)}_{\m_1\cdots\m_{p+1}}}S_{10}=2\k\ls\frac{1}{(p+1)!}C^{(p+1)}_{\m_1\cdots\m_{p+1}}h^{\n\rho}p_{1\,\n}p_{1\,\rho}\right.\\
\left.+\frac{1}{p!}C^{{\rlap{$\mathsurround=0pt\scriptstyle{(p+1)}$}}\hphantom{[\m_1\cdots\m_p}\n}_{[\m_1\cdots\m_p}\lp th_{\m_{p+1}]\n}-h_{\m_{p+1}]}^{\hphantom{\m_{p+1}]}\rho}p_{1\,\rho}p_{2\,\n}-h_{|\n|}^{\hphantom{|\n|}\rho}p_{1\,\m_{p+1}]}p_{1\,\rho}\rp\right.\\
\left.-\frac{1}{(p-1)!}C^{{\rlap{$\mathsurround=0pt\scriptstyle{(p+1)}$}}\hphantom{[\m_1\cdots\m_{p-1}}\n\rho}_{[\m_1\cdots\m_{p-1}}h_{\m_p|\n|}p_{1\,\m_{p+1}]}p_{2\,\rho}\rs.
\end{multline}
Then the amplitude is given by
\begin{multline}
\sqrt{2}\m_p\k^2\e^{a_1\cdots a_{p+1}}\left\{\frac{1}{t}\ls\frac{1}{(p+1)!}C^{(p+1)}_{a_1\cdots a_{p+1}}h^{\m\n}p_{1\,\m}p_{1\,\n}\right.\right.\\
\left.\left.+\frac{1}{p!}C^{{\rlap{$\mathsurround=0pt\scriptstyle{(p+1)}$}}\hphantom{a_1\cdots a_p}\m}_{a_1\cdots a_p}\lp th_{a_{p+1}\m}-h_{a_{p+1}}^{\hphantom{a_{p+1}}\n}p_{1\,\n}p_{2\,\m}-h_\m^{\hphantom{\m}\n}p_{1\,a_{p+1}}p_{1\,\n}\rp\right.\right.\\
\left.\left.-\frac{1}{(p-1)!}C^{{\rlap{$\mathsurround=0pt\scriptstyle{(p+1)}$}}\hphantom{a_1\cdots a_{p-1}}\m\n}_{a_1\cdots a_{p-1}}h_{a_p\m}p_{1\,a_{p+1}}p_{2\,\n}\rs\right.\\
\left.+\frac{2}{s}\ls\frac{1}{(p+1)!}C^{(p+1)}_{a_1\cdots a_{p+1}}\lp rh^b_{\hphantom{b}b}+2h^{bi}p_{1\,b}p_{1\,i}\rp-\frac{1}{p!}C^{{\rlap{$\mathsurround=0pt\scriptstyle{(p+1)}$}}\hphantom{a_1\cdots a_p}i}_{a_1\cdots a_p}\lp h^b_{\hphantom{b}b}p_{1\,a_{p+1}}p_{2\,i}+2h^b_{\hphantom{b}i}p_{1\,a_{p+1}}p_{1\,b}\rp\rs\right\}\\
=\sqrt{2}\m_p\k^2\e^{a_1\cdots a_{p+1}}\left\{\frac{1}{(p+1)!}C^{(p+1)}_{a_1\cdots a_{p+1}}\lp\frac{2r^2}{st}h^b_{\hphantom{b}b}+\frac{4r}{st}h^{bi}p_{1\,b}p_{1\,i}+\frac{1}{t}h^{ij}p_{1\,i}p_{1\,j}\rp\right.\\
\left.+\frac{1}{p!}C^{{\rlap{$\mathsurround=0pt\scriptstyle{(p+1)}$}}\hphantom{a_1\cdots a_p}i}_{a_1\cdots a_p}\lp\frac{r}{t}h_{a_{p+1}i}-\frac{1}{t}h_{a_{p+1}}^{\hphantom{a_{p+1}}j}p_{1\,j}p_{2\,i}-\frac{2r}{st}h^b_{\hphantom{b}b}p_{1\,a_{p+1}}p_{2\,i}\right.\right.\\
\left.\left.-\frac{4r}{st}h^b_{\hphantom{b}i}p_{1\,a_{p+1}}p_{1\,b}-\frac{1}{t}h_i^{\hphantom{i}j}p_{1\,a_{p+1}}p_{1\,j}\rp-\frac{1}{(p-1)!}\frac{1}{t}C^{{\rlap{$\mathsurround=0pt\scriptstyle{(p+1)}$}}\hphantom{a_1\cdots a_{p-1}}ij}_{a_1\cdots a_{p-1}}h_{a_pi}p_{1\,a_{p+1}}p_{2\,j}\right\}.
\end{multline}
By substituting the polarizations for the dilaton or graviton into (\ref{eq:RRSymmetricNSNS}) we can check that both of the field theory results above also agree with the disc amplitude computation.

Since all two-point functions agree between the string and field theory computations (at lowest derivative order), we feel justified in expressing some confidence in the techniques which we have outlined in this paper.

\bibliographystyle{utphys}

\providecommand{\href}[2]{#2}\begingroup\raggedright\endgroup


\end{document}